\documentclass[11pt,a4paper]{article}

\usepackage[utf8]{inputenc}    
\usepackage[T1]{fontenc}       
\usepackage{lmodern}           
\usepackage{geometry}          
\usepackage{graphicx}          
\usepackage{amsmath, amssymb}  
\usepackage{physics}           
\usepackage{hyperref}          
\usepackage{siunitx}           
\usepackage{float}             
\usepackage{caption}           
\usepackage{subcaption}        
\usepackage{booktabs}          
\usepackage[numbers,sort&compress]{natbib}            
\usepackage{doi}               
\usepackage{xcolor}            
\usepackage{authblk}           
\usepackage{microtype}         
\usepackage{enumitem}          
\usepackage{url}               
\usepackage{cancel}               
\usepackage[font=footnotesize]{caption}
\usepackage{comment}
\usepackage{bbding}
\usepackage{endnotes}
\let\footnote=\endnote

\title{The Theory of Economic Complexity}
\author[1,2,3]{C\'esar A. Hidalgo\thanks{Corresponding author: cesar.hidalgo@tse-fr.eu}}
\author[1,2,4]{Viktor Stojkoski}
\affil[1]{\small Center for Collective Learning, IAST, Toulouse School of Economics}
\affil[2]{\small Center for Collective Learning, CIAS, Corvinus University of Budapest}
\affil[3]{\small Alliance Manchester Business School, University of Manchester}
\affil[4]{\small Ss. Cyril and Methodius University in Skopje}
\date{}

\begin{document}

\maketitle

\begin{abstract}
We provide a mechanistic foundation for economic complexity methods. In our model, an economy’s ability to produce an activity depends on the joint presence of required capabilities. We analytically derive the Economic Complexity Index (ECI) for this model and show that it is a monotonic function of the stock of capabilities in an economy. We then explore the family of functions and conditions that are compatible with economic complexity estimates and show that multiplicatively separable production functions are incompatible with economic complexity estimates. By contrast, additive non-separable functions are sufficient regardless of whether they are log super-modular. We also show that this model explains differences in the shape of networks of related activities, such as the product space or research space. These findings solve long standing puzzles in the literature on economic complexity.
\end{abstract}

\section{Introduction}
\label{sec:introduction}

A key tenet of the economic complexity literature is the idea that the combined presence of the capabilities or factors needed for production can be estimated without having to define them. This is an instance of a more general inference problem in complex-systems, where one observes a bipartite pattern of participation (in economic complexity, a network connecting economies to activities) and attempts to infer a low-dimensional representation to explain the structure and dynamics of the system~\cite{dhillon2001co,boccaletti2006complex,von2007tutorial,latapy2008basic}. This setup is central to two contributions that ignited the study of economic complexity at the beginning of this century: measures of relatedness and complexity.\\

Relatedness measures--which are estimates of the affinity or potential between an economy and an activity--where introduced together with the product space~\cite{hidalgo_product_2007}, a network connecting products based on the idea that ``if two goods are related because they require similar institutions, infrastructure, physical factors, technology, or some combination thereof, they will tend to be produced in tandem.'' This network is used to create estimates of economic potential that do not rely on defining specific factors or capabilities, but that instead leverage implicit information about shared capabilities based on patterns of co-specialization. These and other networks of related activities, such as the product space~\cite{hidalgo_product_2007,hausmann_atlas_2014}, the industry space~\cite{neffke_how_2011,neffke_skill_2013,jara-figueroa_role_2018}, the research space~\cite{guevara_research_2016,chinazzi_mapping_2019}, and the technology space~\cite{kogler_mapping_2013,kogler_mapping_2015}, have become important tools in economic geography, innovation studies, and international development, as they provide a means to formalize notions of path dependency and estimate the chances that an economy has developed or accumulated the factors or capabilities needed to engage in an activity.\\

Metrics of economic complexity where introduced soon after (~\cite{hidalgo_building_2009}) in an attempt to provide an outcomes based estimate of the combined presence of the factors or capabilities available in an economy. These metrics of complexity have also become useful tools in economic geography, international development, and innovation, because of their ability to explain international and regional variations in economic growth~\cite{hidalgo_building_2009,hausmann_atlas_2014,chavez_economic_2017,domini_patterns_2022,koch_economic_2021,stojkoski_impact_2016,stojkoski_relationship_2017,ourens_can_2012,poncet_economic_2013,stojkoski_multidimensional_2022,tacchella_new_2012,bustos_production_2022,atkin2021globalization,teixeira2022economic,hoeriyah2022economic,mao2021economic,basile_economic_2022,cardoso2023export,romero2021economic,perez2019measuring}, inequality~\cite{hartmann_linking_2017,bandeira_morais_economic_2018,sbardella_economic_2017,le_caous_economic_2020,ben_saad_economic_2019,fawaz_spatial_2019,zhu_export_2020,barza_knowledge_2020,lee2021economic}, and sustainability outcomes~\cite{can_effects_2017,neagu_link_2019,neagu2019relationship,romero_economic_2021,lapatinas_economic_2019,mealy_economic_2020,gomez2022income,balsalobre2023influence,silveira2025economic,dordmond_complexity_2020}.\footnote{Among other outcomes~\cite{guneri_does_nodate,vu_does_2020,ferraz2018economic,lapatinas_eu_2023,barza_cities_2024,yenilmez2025understanding,djeunankan2025hidden,lizardo_mutual_2018}.}. Yet, despite several attempts to develop a mathematical theory of economic complexity~\cite{hidalgo_building_2009,hausmann_atlas_2014,hausmann_network_2011,atkin2021globalization,bustos_production_2022,schetter2022measure,mealy_interpreting_2019,yildirim_sorting_2021,cakir2021amalgamation,servedio_economic_2024,bottai_reinterpreting_2024,mcnerney_bridging_2023,utkovski_economic_2018,desmarchelier2018product}, we still lack an analytical connection between the metrics used in the empirical literature and a production function based model from where we can derive these metrics from first principles. \footnote{For a review of the field see \cite{hidalgo_economic_2021,balland_new_2022}.} More to the point, we lack a theory that we can map one-to-one to the many observables used in the empirical literature, such as specialization matrices, binary specialization matrices, and eigenvectors. This paper provides an analytical model that generates the most widely user observables in empirical estimations of economic complexity and relatedness. \\ 

Here, we connect the empirical economic complexity literature with a theoretical model to provide four contributions.\\

First, we provide an analytical derivation of the eigenvector known as the economic complexity index or $ECI$. This is a model where economies (such as countries or cities) differ in the capabilities required by each activity (such as products or industries) and where the output of an economy is constrained by the capabilities it has developed, while the geography of an activity is limited to the places that have developed the capabilities the activity requires. We solve the one capability instance of this model analytically and show that the economic complexity index, or $ECI$, is a vector separating economies with an above average capability level from those with a below average capability level. Interestingly, this property is independent of how capabilities are distributed and can be generalized to other production functions. We find that separable production functions are incompatible with economic complexity estimates and show that simple non-separable functions are sufficient for economic complexity estimates to work.\\

Second, we extend this result numerically to models involving many capabilities, including heterogeneous capabilities assigned idiosyncratically to each economy. We show that in this case $ECI$ is a monotonic function of the average capability level of an economy and recovers the first singular vector of the capability matrix. By exploring models combining correlated and uncorrelated capabilities, we show this result to be robust to substantial levels of noise, holding even when more than 50 percent of the capabilities are assigned at random. This helps show that $ECI$ is a measure of economic complexity, as it captures whether an economy has multiple capabilities without having to make assumptions about their nature.\\

Third, we extend the single capability model to a short-run equilibrium framework where we calculate wages, prices, and consumption. We show analytically that under these assumptions $ECI$ still separates economies among those with high and low capability levels. We also determine an equilibrium wage to help interpret the well-established empirical relationship between economic complexity and growth, and show that the prices of goods in this model follows a convex function of their capability requirements, indicating a higher premium for complex goods.\\

Finally, we use the multi-capability model to explain known variations in the shapes of networks of related activities, such as the product space (based on product co-exports) and the research space (based on co-publication patterns). We show  that the core-periphery structure observed in the product space~\cite{hidalgo_product_2007}, comes from correlated capability levels and that the ring structures observed in networks of related research fields~\cite{guevara_research_2016,borner_design_2012} can be explained by capability levels following a Toeplitz circulant matrix.\\

There are a few reasons why these results should be of interest to a wide interdisciplinary audience, including scholars working at the intersection of complex systems, economic complexity, and economic development.\\

First, while the economic complexity index or $ECI$ enjoys wide adoption in policy circles\footnote{For example, it is the number one mission of Malaysia's New Industrial Master Plan~\cite{nimp2030}, it was used in the recent European competitiveness report by Mario Draghi~\cite{draghi2024future}, and it is a key development target for rich resource intensive economies, such as Saudi Arabia and the United Arab Emirates. It has also motivated the creation of regional reports for Australia~\cite{reynolds_sub-national_2018}, Turkey~\cite{erkan_economic_2015}, Uruguay~\cite{ferreira-coimbra_evolucion_2009}, Russia~\cite{lyubimov_atlas_2018,lyubimov_economic_2017}, Mexico~\cite{zaldivar_economic_2019,perez_hernandez_diagnostico_2019}, Quebec~\cite{wang_economic_2020}, and Italy~\cite{basile_economic_2019}, among other places.}, the lack of a theoretical foundation has left it open to criticism of being an ad-hoc or uninterpretable measure~\cite{valverde-carbonell_rethinking_2025,atkin2021globalization,sciarra_reconciling_2020,mcnerney_bridging_2023}. Our findings provide a clear interpretation for $ECI$ in the context of multi-factor model of production. We show that $ECI$ is a monotonic estimate of the capability stock of an economy derived from a multi-product specialization matrix--an interpretation that is consistent with previous work exploring the interpretability of the $ECI$ as a clustering method~\cite{mealy_interpreting_2019,bottai_reinterpreting_2024,servedio_economic_2024} and connecting $ECI$ with the notion of log-supermodularity~\cite{schetter2022measure}.\\

Second, these findings dispel the notion that economic complexity is a measure of generalized diversity, as it was originally suggested~\cite{hidalgo_building_2009}. The analytical solutions show that economies specialized in the largest number of activities (the more diverse economies) are not necessarily the ones with the highest capability stocks\footnote{The notion that economic complexity is different from diversity was noted theoretically by~\cite{mealy_interpreting_2019} and has been in the literature from early on, since the work introducing $ECI$ showed that measures of diversity or concentration, such as entropy or the Herfindahl-Hirschmann index (HHI) failed to explain future economic growth as $ECI$ did~\cite{hidalgo_building_2009}.}. In fact, the model predicts that economic development is a process of diversification only until a certain point, since economies with the highest capability stocks are expected to specialize in complex activities--and are therefore--less diverse than slightly less complex economies. This provides a theoretical foundation for the finding that countries at high-levels of development tend to specialize (e.g., Imbs and Wacziarg~\cite{imbs_stages_2003}) and is consistent with the notion that $ECI$ is higher for ``small,'' sophisticated and somewhat specialized economies, such as those of Singapore, Switzerland, and Finland\footnote{While larger and more diverse economies, like those of Spain and Italy, are not necessarily as complex.}. Still, the model predicts a positive correlation between capability levels and diversity, but through a non-monotonic function, explaining why measures of diversity or concentration are non-ideal estimates of the complexity of an economy.\\  

Third, these results also provide a means to interpret the structure of the networks of related activities, such as the product space~\cite{hidalgo_product_2007}, industry space~\cite{neffke_how_2011}, or research space~\cite{guevara_research_2016}. These networks have been used extensively to model path dependencies and generate measures of export or employment potential~\cite{hidalgo_principle_2018,kogler_mapping_2015,guevara_research_2016,neffke_how_2011,jara-figueroa_role_2018,neffke_skill_2013,jun_bilateral_2019,hidalgo_economic_2021,farinha_what_2019,balland2016dynamics,boschma_emergence_2013,farinha_what_2019-1,chen_inter-industry_2017,ferrarini_product_2015,alabdulkareem_unpacking_2018}. Yet, the structure of these networks differs depending on the data used to generate them. For instance, networks derived from co-export data, are known to have a dense core of complex products surrounded by a periphery of low complexity products ~\cite{hidalgo_product_2007}. Networks connecting research fields, by citations~\cite{borner_design_2012} or co-authorships~\cite{guevara_research_2016}, follow a ring structure, where fields are connected to a few neighbors and the network lacks a clear center.\footnote{In simple, the ring is: medicine, biology, chemistry, physics, computer science and math, economics, cognitive science, neuroscience, and back to medicine.}. While these differences in structure are self-evident, we hitherto lacked a mechanistic model to explain them. Here, we show how to generate network structures that resemble those observed in the empirical literature by changing the shape of the capability matrices.\\

Finally, our analysis contributes to a broader complex systems literature investigating how latent factors shape emergent low-dimensional structure, by clarifying how a simple generative model with unobserved, heterogeneous, and complementary constraints can produce the network and spectral regularities extracted from output data. Indeed, the economic complexity toolkit has been used in domains beyond economic development, including ecology~\cite{baudena2015revealing,dominguez2015ranking}, health~\cite{garas2021development,cross2025complexity}, scientometrics ~\cite{guevara_research_2016,yenilmez2025understanding}, etc., suggesting that the theoretical connections established here can help interpret analogous projections and spectral summaries in these settings as well.\\

Together, these findings enrich the theoretical foundations of economic complexity and position its core tools and results as a useful lens for other, similar complex-systems settings.\\

\subsection{Empirical and Theoretical Work in Economic Complexity}

Empirical work in economic complexity usually starts with matrices summarizing the geography of many economic activities (e.g., exports by country and product, payroll by city and industry, patents by city and technology, etc.). These rectangular matrices (or bipartite networks) are then used for two things. The first one is to estimate networks of similar activities~\cite{kogler_mapping_2013,kogler_mapping_2015,guevara_research_2016,hidalgo_product_2007,hidalgo_principle_2018,hidalgo_economic_2021,neffke_how_2011,ma2025disparities,muneepeerakul2013urban,alabdulkareem_unpacking_2018,chinazzi_mapping_2019} or similar economies
\cite{bahar_neighbors_2014}, which are then used to estimate the ``relatedness'' or affinity between an economy and an activity. These estimate are then used to estimate which economies are more likely to enter (and less likely to exit) activities that share capabilities with each other\footnote{What is know in the specialized literature as The ``Principle of Relatedness''~\cite{hidalgo_principle_2018}.}.\\

The second one is to estimate the value of the portfolio of activities an economy specializes in, known as measures of economic complexity~\cite{hidalgo_building_2009,hidalgo_economic_2021,tacchella_new_2012,cristelli_measuring_2013,hausmann_atlas_2014,bustos_production_2022,atkin2021globalization,sciarra_reconciling_2020-1}. These measures were also motivated as agnostic estimates of the capabilities available in an economy~\cite{hidalgo_building_2009} and are often based on the assumption that high-complexity economies specialize in high-complexity activities. In fact, the economic complexity index or $ECI$, defines the complexity of an economy as the average complexity of the activities it specializes in, and the complexity of an activity as the average complexity of the economies specialized in that activity.\footnote{A similar definition was proposed over a decade later by~\citep{atkin2021globalization}. In their words: ``If a country is known to be more capable than another, say the United States (US) versus Bangladesh (BG), then one can identify any good $k$ as more complex than another reference good $k_0$ if, relative to the reference good, it is more likely
to be exported by the United States than Bangladesh. [$\dots$] Conversely, if a good is known to be more complex than another, say medicines (ME) versus men’s underwear (UW), then one can identify any country $i_1$ as more capable than another reference country $i_0$ if, relative to the reference country, it is more likely to export medicines than underwear. ''}.\\

These measures of complexity, in particular $ECI$, enjoy wide adoption in international and regional development circles, as they have been shown to be robust estimators of future economic growth~\cite{hidalgo_building_2009,hausmann_atlas_2014,chavez_economic_2017,domini_patterns_2022,koch_economic_2021,stojkoski_impact_2016,stojkoski_relationship_2017,ourens_can_2012,poncet_economic_2013,stojkoski_multidimensional_2022,tacchella_new_2012,bustos_production_2022,atkin2021globalization}, and of international variations in inequality~\cite{hartmann_linking_2017,bandeira_morais_economic_2018,sbardella_economic_2017,le_caous_economic_2020,ben_saad_economic_2019,lee2021economic,gomez2022income}, and emissions~\cite{can_effects_2017,neagu_link_2019,neagu2019relationship,romero_economic_2021,lapatinas_economic_2019,mealy_economic_2020,caldarola2024economic,balsalobre2023influence,silveira2025economic}.\\

These two strands of literature support the notion that an economy's pattern of specialization matters for subsequent economic development~\cite{hausmann_what_2007,rodrik_whats_2006}, which has been a key intuition motivating these efforts\footnote{This policy intuition is connected to an old debate in development economics, going back to at least Alexander Hamilton's Report on Manufactures~\cite{hamilton_report_1791}, which advocated for the industrialization of the United States, and has been central to the works of scholars such as Rosenstein-Rodan~\cite{rosenstein-rodan_notes_1961,rosenstein-rodan_problems_1943}, Rostow\cite{rostow_stages_1959}, Hirschman\cite{hirschman_generalized_1977}, Prebisch\cite{prebisch_economic_1962}, Gerschenkron\cite{gerschenkron_early_1963}, and Balassa\cite{balassa_exports_1985}. For a discussion on how these different development theories related to economic complexity see~\cite{hidalgo_policy_2023}.}. Yet, despite copious empirical work, we still lack an understanding of why the eigenvectors used as measures of economic complexity are good predictors of an economy's subsequent growth and development.\\

Theoretical work on economic complexity has focused instead on the construction of models of development and innovation that either i) follow a combinatorial tradition ~\cite{weitzman_recombinant_1998,kauffman_origins_1993,hidalgo_building_2009,kremer_o-ring_1993,hausmann_network_2011,tria_dynamics_2014,cristelli_measuring_2013,fink_how_2019,pichler2020technological,mcnerney2011role,fleming2001recombinant,o2021productive,van2022variety} or ii) assume a log-supermodular productivity assignment~\cite{costinot2009origins,schetter2022measure,yildirim_sorting_2021}. \\

The first group of models build on the notion that economies develop capabilities~\cite{hidalgo_building_2009,hausmann_network_2011}, defined following notions common in the evolutionary economics literature ~\cite{dosi1982technological,teece1997dynamic,lall1992technological,nelson_evolutionary_1982}. Here capabilities are the productive knowledge and skills that enable economic activity, and activities differ in the combinations of the capabilities they require. Since these capabilities are complementary, producing an activity requires the simultaneous presence of many of them. That is why these theories have been dubbed as the ``Lego'' or ``Scrabble'' theories of development. In these models, the ability of economies to produce a product depends on having the right combination of capabilities, like in a proverbial game of scrabble where products are ``words'' and economies have a stock of ``letters.''\\

The second group can be viewed as a tractable extension of the capabilities tradition introduced in~\cite{hidalgo_building_2009,hausmann_network_2011} that imposes additional structure on heterogeneity. These models summarize the relevant heterogeneity with one-dimensional types: a scalar attribute of places (e.g., productivity, or an aggregate capability level) and a scalar attribute of activities (e.g., intensity or complexity). Under log-supermodularity\footnote{Formally, a production function $f(c,p)$ is log-supermodular if, for any two economies $c^{'} > c$ and any two activities $p^{'} > p$ $f_{c'p'}/f_{c'p}>f_{cp'}/f_{cp}$}, this structure delivers positive assortative matching meaning that high-type places optimally sort into high-intensity activities. This further implies a sharp ordering in equilibrium production patterns, providing micro-foundations for the observed form of real world matrices including economic complexity metrics~\cite{schetter2022measure,yildirim_sorting_2021}.\\

Our approach adds to these traditions. Rather than imposing a one-dimensional ordering as a sufficient condition for sorting, we derive economic complexity metrics from a generative model in which production depends on combinations of multiple, partially independent capabilities. This distinction is consequential because economic complexity is not computed from production levels or intensities directly, but from specialization patterns that are normalized and discretized. As we show, log-supermodularity at the level of production is not necessary for these transformed objects to preserve information about underlying capability levels: even technologies that generate sharp sorting in outputs can become uninformative after normalization.\\

More importantly, the one-dimensional structure implied by log-supermodularity is at odds with key empirical features of some of the observed networks. For instance, the research space exhibits a ring structure, and the product space displays well-defined clusters, both of which reflect a multidimensional capability structure in which relatedness depends on which capabilities are shared, not merely on their aggregate level. These topological features cannot be captured by a single latent index, even when such an index provides a useful summary for ranking purposes. By explicitly modeling capabilities as a vector and tracing how specialization patterns emerge from their combinations, our framework explains simultaneously why a global ranking such as the ECI can be informative, and why the observable geometry of knowledge and production networks departs from a one-dimensional ordering.\\

In particular, our approach builds on the combinatorial model introduced by~\cite{hidalgo_building_2009}, which is a generalization of Kremer's O-Ring model of development~\cite{kremer_o-ring_1993} or weak-link production functions\cite{jones2013ring}. More precisely, we call this the Kremer-Shockley model of productivity, since the same multiplicative formula was introduced in 1957 by William Shockley in a paper explaining productivity differences among researchers~\cite{shockley1957statistics}.\\

The Kremer-Shockley model assumes a multi-step production process where the output of an economy is the product of the probabilities that it succeeds at each step. In other words, producing an item in this model requires a sequence of tasks, each of which has a probability of failing. This implies that the output of an economy decays exponentially with the length of the production chain at a rate determined by the probability of succeeding at a task. The key outcome is that economies with higher probabilities of completing a task should specialize in activities requiring multiple steps.\footnote{See also~\cite{melitz2014missing} for an extension of the O-Ring model to trade.}\\

We study a generalized version of this framework in which economies have probabilities of possessing capabilities and activities differ in the probabilities with which they require them. This preserves the complementarity structure of the Kremer-Shockley model while allowing for an arbitrary number of economies, activities, and capabilities, all of which may be heterogeneous. Because production is multiplicative in these probabilities, the implied production mapping is log-supermodular by construction, generating a systematic ordering whereby economies with higher capability levels have relatively higher output in activities with higher capability requirements. This ordering provides the structural link between the combinatorial capabilities tradition and the sorting logic emphasized in log-supermodular assignment models~\cite{schetter2022measure}.\\

Throughout the paper we use production functions in a limited and formal sense 
that differs from the aggregate concept subject to well-known critiques~\cite{felipe2013aggregate, 
nelson_evolutionary_1982,dosi1982technological}. We do not map aggregate capital and labor into aggregate output. Instead, we use production functions as generative mappings from latent capability profiles to observable patterns. That is, the observables used in economic complexity estimates, are not production levels but specialization patterns that need to be constructed from output data through normalization and discretization. The key question we address is whether the low-dimensional structure implied by the underlying production mapping survives these transformations.\\

We find that, for a wide variety of model specifications, $ECI$ recovers the probability that an economy has multiple capabilities, even when these are highly noisy or idiosyncratic. We then generalize this finding to a Cobb-Douglas type factor intensity production function and find that the ability of $ECI$ to separate among economies with higher and lower capability levels can be generalized to any shifted production function of the form $Y_{cp}=B_c+f_cg_p$ where $f_c$ is a general function characterizing an economy $c$, $g_p$ is a general function characterizing an activity $p$, and $B_c$ is an economy specific term that represents a baseline output level.\\

As in Kremer-Shockley~\cite{kremer_o-ring_1993,shockley1957statistics}, we start from a supply-side model that assumes prices are exogenous and do not provide an explicit model of wages or demand. So, we then embed the single capability model in a short-run equilibrium framework to estimate functions for the implied wages, consumption, and prices. We show that wages increase with the capability level of an economy, consumption grows with income, and prices are higher for more complex products (products that are more demanding of the capability). This also allows us to show that our main result--that the economic complexity eigenvector separates high- from low-capability economies--holds after introducing these additional assumptions.\\

Finally, we use this model to explain the structures of networks of related activities, such as the product space and research space, and show that it is possible to generate networks with a similar structure than the ones observed in the empirical literature by manipulating the capability and capability requirement matrices.\\

The remainder of the paper is organized as follows. The next section (Section~\ref{sec:methods}) introduces the single-capability model and presents an analytical derivation of the economic complexity index ($ECI$). Section~\ref{sec:multi_capability-model} generalizes these results numerically to several versions of a multi-capability model. Section~\ref{sec:other_production_functions} explores additional production functions and Section~\ref{sec:prices-wages-consumption} embeds the model in a short-run equilibrium framework. Section~\ref{sec:relatedness} uses the model to explain the structure of networks of related activities, and Section~\ref{sec:conclusion} concludes.\\

\section{The Single Capability Model}
\label{sec:methods}

Our model assumes that an economy $c$ has capability $b$ with probability $r_{c,b}$ and that activity $p$ requires capability $b$ with probability $q_{p,b}$. Here, $r_{c,b}$ captures the presence of capability $b$ in economy $c$. That is, the probability that economy $c$ possesses the productive input that capability $b$ represents, such as a skill, technology, institution, or piece of infrastructure. Similarly, $q_{p,b}$ captures the factor intensity of activity $p$ with respect to capability $b$, i.e., the degree to which activity $p$ relies on that input. Activities with higher factor intensity place greater demands on the capabilities of the economies that produce them.

Capabilities should be understood as a non-fungible input. For example, a form of human capital such as specialized skills or knowledge that economies accumulate and that directly enable production. Yet, the model is agnostic about the nature of capabilities (e.g. they could also include specialized institutions or norms). That is, we treat these capabilities as nuanced and dynamic factors of production that economies hold in varying degrees. We acknowledge that this model abstracts from other traditional factors of production such as raw labor and physical capital, focusing instead on capabilities as the key source of variation across economies and activities. This abstraction is deliberate since our goal is not to provide a complete account of all factors of production, but to show that the pattern of capability accumulation alone is sufficient to explain the structure of the empirical results in economic complexity.

For pedagogical reasons we start with a single capability or factor and an arbitrary number of economies and activities (that is $r_{c,b}\xrightarrow{}r_c$ and $q_{p,b}\xrightarrow{}q_p$). This case will allow us to get a basic intuition that we will then generalize to more complex versions of the model. The advantage of starting with the single capability model is that we can derive its economic complexity index ($ECI$) analytically.\\

Let the output $Y_{cp}$ of economy $c$ in activity $p$ be given by the matrix:\\
\begin{equation}
    Y_{cp}=A(1-q_p(1-r_c)),
    \label{one_capability_model}
\end{equation}\\
where $A$ is a constant or scale factor and $1-q_p(1-r_c)$ is the probability that economy $c$ has the capability that product $p$ requires. This probability is written as a complement. That is, one minus the probability that the activity requires a capability ($q_p$) that the economy does not have ($1-r_c$).\footnote{We later explore other forms including a search term, such as: $  Y_{cp}=r_c^\alpha(1-q_p(1-r_c))$ where $r_c^\alpha$ models how difficult it is to find a capability in an economy and the $1-q_p(1-r_c)$ models how good of a fit it is.} In matrix form, the output matrix implied by this model is given by:\\
\begin{equation}
Y_{cp}=A \begin{bmatrix}
1-q_1(1-r_1) & 1-q_2(1-r_1) & \dots \\
1-q_1(1-r_2) & \dots & \dots \\
\dots & \dots & 1-q_N(1-r_N)
\end{bmatrix}.
\label{eq:one_capability_formula}
\end{equation}\\

Going forward, we sort rows in descending order of $r$ and columns in ascending order of $q$. That is, the first cell of the matrix ($Y_{11}$) is the output of the economy with the highest probability of having the capability in the activity with the lowest probability of requiring it. This sorting convention will greatly facilitate the visual inspection of these matrices.\\

This sorting implies an ordered structure in which economies with higher probabilities of possessing the capability produce relatively more in activities that are more demanding of that capability. When rows are ordered by $r_c$ and columns by $q_p$, this structure induces a monotone pattern in $Y_{cp}$ that satisfies ``log-supermodularity.'' This property is shared with a class of assignment-based models that generate ordered production patterns through complementarity, but here it arises directly from the probabilistic capability formulation~\cite{schetter2019structural,yildirim2021sorting}. Formally, a production function $f(c,p)$ is log-supermodular if, for any two economies $c^{'} > c$ and any two activities $p^{'} > p$:
\begin{equation}
    \frac{f(c^{'}, p^{'})}{f(c^{'}, p)} > \frac{f(c, p^{'})}{f(c, p)}.
    \label{eq:log_supermodularity}
\end{equation}\\

For differentiable production functions, it has a continuous analog: $f(r_c,q_p)$ is log-supermodular if and only if $\frac{\partial^2 \ln f}{\partial r_c\, \partial q_p} > 0$. Intuitively, this condition captures complementarity between presence of factors and factor intensities. This means that economies with higher capability levels gain relatively more from engaging in more demanding activities, generating the systematic sorting pattern that underlies our analysis.\\

A key difference between this implementation of the model and previous work~\cite{hidalgo_building_2009,hausmann_network_2011,cristelli_measuring_2013} is that here we use the model to generate an output matrix $Y_{cp}$, whereas earlier studies typically applied the model directly to simulate specialization matrices. In practice, however, the objects used to compute $ECI$ are not production levels but specialization indicators constructed after normalizing and discretizing raw output or trade data. These transformations are designed to remove scale effects and isolate relative specialization, and they need not preserve the ordered or supermodular structure present in the underlying production mapping.\\
 
(i) Estimating the matrix of revealed comparative advantage or RCA $R_{cp}$ according to Balassa's (1965) definition~\cite{balassa_trade_1965}. This matrix normalizes the output matrix $Y_{cp}$ by the sum of its rows and columns and it is equivalent to a matrix comparing the observed output ($Y_{cp}$) (see eqn.~\eqref{rca}). This step normalizes $Y_{cp}$ for differences in economy size and activity market size so that entries capture relative specialization rather than scale, making economies and activities comparable in the cross-section.\\
 
(ii) Estimating the binary specialization matrix $M_{cp}$. This is a matrix that is 1 if $R_{cp}\geq1$ and 0 otherwise. This binary matrix is motivated in the empirical literature as a means to remove the tails of the $R_{cp}$ matrix, since the ratio definition of $R_{cp}$ results in larger variance for economies with low levels of output (small $Y_c=\sum_p Y_{cp}$) and activities with small markets (small $Y_p=\sum_c Y_{cp}$).\\
 
(iii) Estimating the complexity matrix $M_{cc'}$. This is a square matrix connecting economies with similar specialization patterns and is the one used to derive the economic complexity index. This matrix is defined using the reciprocal averaging method known as the method of reflections~\cite{hidalgo_building_2009}, but it can also be defined as the product of four matrices (we will introduce the exact formula at that point).\\

We begin with the standard definition of the RCA matrix or $R_{cp}$ which is:\\

\begin{equation}
    R_{cp}=\frac{Y_{cp}\sum_{c,p}Y_{cp}}{\sum_{c}Y_{cp}\sum_p Y_{cp}}.
    \label{rca}
\end{equation}\\

Also, since it will simplify the math going forward, we use Einstein's notation, where summed indices are ``suppressed'' or ``muted'' (e.g. $Y_c=\sum_{p}Y_{cp}$). In this notation $R_{cp}$ takes the more compact form:

\begin{equation}
    R_{cp}=\frac{Y_{cp}Y}{Y_{c}Y_{p}}.
\end{equation}\\

To estimate $R_{cp}$ for the single capability model we need to notice a couple of things. First, since the scale factor $A$ is common to all terms, it cancels out of $R_{cp}$ (so we can ignore it). Second, we should notice that applying the sum operator to the terms in $R_{cp}$ transforms variables into averages. We can illustrate this by using the sum over $p$ as an example (the derivation is analogous for the other terms):\\
\begin{equation}
\begin{aligned}
    Y_p&= \sum_{p}(1-q_p(1-r_c)),\\
    Y_p&=N_p-(1-r_c)\sum_p{q_p},\\
    Y_p&=N_p(1-(1-r_c)\langle q \rangle),
\end{aligned}
\end{equation}
where $N_p$ is the number of activities or products and $\langle q \rangle$ is the average of $q_p$ over all activities. Using this property, we can now rewrite $R_{cp}$ as:\\

\begin{equation}
    R_{cp}=\frac{(1-q_p(1-r_c))(1-\langle q \rangle(1-\langle r \rangle))}{(1-q_p(1-\langle r \rangle)(1-\langle q \rangle (1-r_c))}.
\end{equation}

To derive $M_{cp}$ we need to identify when $R_{cp}$ is larger or smaller than one. We can do this by manipulating the inequality.\\
\begin{equation}
    (1-q_p(1-r_c))(1-\langle q \rangle(1-\langle r \rangle))\geq(1-q_p(1-\langle r \rangle)(1-\langle q \rangle (1-r_c)).
\end{equation}\\
This simplifies to:
\begin{equation}
    q_p(1-r_c)+\langle q \rangle(1-\langle r \rangle) \leq
    q_p(1-\langle r \rangle)+\langle q \rangle(1-r_c), 
\end{equation}\\
leading to the condition:
\begin{equation}
   (r_c-\langle r \rangle)(q_p-\langle q \rangle) \geq 0.
\end{equation}
Since this is an inequality, we need to be careful about the signs of $(q_p-\langle q \rangle)$ and $(r_c-\langle r \rangle)$. Changes in sign flip the inequality operator. So what this condition means is that $R_{cp}\geq1$ when $r_c \geq \langle r \rangle$ and $q_p-\langle q \rangle \geq 0$ or when $r_c < \langle r \rangle$ for $q_p-\langle q \rangle < 0$. We can also get this condition intuitively by considering the case when $q_p=\langle q \rangle$ or $r_c=\langle r \rangle$. In these two cases $R_{cp}=1$, meaning that these lines divide the matrix into regions where the values of $R_{cp}$ are higher or smaller than one. In sum, from the condition above $M_{cp}$ is a matrix divided into four quadrants:\\

\begin{equation}
\begin{aligned}
    M_{cp}&=1, \quad \text{if} \quad r_c \geq \langle r \rangle \quad \& \quad q_p \geq \langle q \rangle, \quad\\
     M_{cp}&=1, \quad \text{if} \quad r_c < \langle r \rangle \quad \& \quad q_p < \langle q \rangle, \quad\\
    M_{cp}&=0, \quad \text{if} \quad r_c < \langle r \rangle \quad \& \quad q_p \geq \langle q \rangle, \quad\\
    M_{cp}&=0, \quad \text{if} \quad r_c \geq \langle r \rangle \quad \& \quad q_p < \langle q \rangle. \quad
\end{aligned}
\end{equation}\\

This matrix represents a world where countries with a high probability of having the capability ($r_c$ higher than average), specialize in products with high probability of requiring the capability ($q_p$ higher than average), and countries with low probability of having the capability specialize in products with low probability of requiring it. This is related to the idea of log-supermodularity in trade theory~\cite{schetter2022measure}.\\

As an example, consider a world with four countries and six products, where two countries have above average $r_c$ and three products have above average $q_p$. In this example, the binary specialization matrix $M_{cp}$ takes the form:\\

\begin{equation}
M_{cp}= \begin{bmatrix}
0 & 0 & 0 & 1 & 1 & 1 \\
0 & 0 & 0 & 1 & 1 & 1 \\
1 & 1 & 1 & 0 & 0 & 0 \\
1 & 1 & 1 & 0 & 0 & 0
\end{bmatrix}.
\end{equation}\\

Finally, we use $M_{cp}$ to derive $M_{cc'}$. Here we use the standard reciprocal average method, or method of reflections. This method proposes that the complexity of an economy is the average complexity of the activities that economy is specialized in, and that the complexity of an activity is the average complexity of the economies specialized in that activity. Using the economic complexity index ($ECI$) and the product complexity index ($PCI$) to indicate the complexity of economies and activities we obtain:\\

\begin{equation}
\begin{aligned}
ECI_c = \frac{1}{M_c}\sum_{p} M_{cp} PCI_p, \\
PCI_p = \frac{1}{M_p}\sum_{c} M_{cp} ECI_c.
\label{eq:eci_system}
\end{aligned}
\end{equation}

Putting the second equation into the first, one can show that $ECI_c$ is the solution to the following self-consistent equation:\\
\begin{equation}
ECI_c = \sum_{c'} M_{cc'} ECI_{c'},
\end{equation}
with
\begin{equation}
M_{cc'}=\frac{1}{M_{c}}\sum_{p}\frac{M_{cp}M_{c'p}}{M_{p}},
\label{Mcc'}
\end{equation}

This means that the economic complexity vector $ECI_c$ must be an eigenvector of the $M_{cc'}$ matrix representing the steady state of the mapping defined by the system in eqns.~\eqref{eq:eci_system} (the same derivation can be used to define the $M_{pp'}$ matrix used to estimate $PCI$).\footnote{$M_{cc'}$ can also be defined as the product of four matrices $M_{cc'}=D_cM_{cp}D_pM_{pc'}$ where $D_c$ is a diagonal matrix of $1/M_c$ and $D_p$ is a diagonal matrix of $1/M_p$.}\\

Estimating the first eigenvector of $M_{cc'}$ is trivial because $M_{cc'}$ is a stochastic matrix (each row adds to one). That means its first eigenvector will always be the vector $\mathbf{1}$. This is easy to prove by summing $M_{cc'}$ over $c'$:\\

\begin{equation}
\begin{aligned}
    M_{cc'}\mathbf{1}&=\sum_{c'}\frac{1}{M_{c}}\sum_{p}\frac{M_{cp}M_{c'p}}{M_{p}},\\
    M_{cc'}\mathbf{1}&=\frac{1}{M_{c}}\sum_{p}\frac{M_{cp}M_{p}}{M_{p}},\\
    M_{cc'}\mathbf{1}&=\frac{1}{M_{c}}\sum_{p}M_{cp}=\mathbf{1}.
\end{aligned}
\end{equation}

Since the first eigenvector is $\mathbf{1}$, the steady state of the system represented by eqns~\eqref{eq:eci_system} is given by the second eigenvector. To estimate that eigenvector, we need to calculate $M_{cc'}$. Here, we consider three cases. When the number of economies and activities is even, when the number of economies is odd and the number of activities is even, and when both the number of economies and activities are odd. The need to consider these cases separately will become self-evident once they are introduced. \\

We begin with the simplest case, that of an even number of economies and activities, which provides the core analytical result showing that ECI separates economies into those with above and below average capability levels. \\

Let $\langle r \rangle$ and $\langle q \rangle$ be the medians of their distributions. In that case, $M_{cc'}$ reduces to a block diagonal matrix with two blocks with values of $1/M_p$ (all economies have the same diversity and all activities the same ubiquity). That is: \\

\begin{equation}
\begin{aligned}
M_{cc'} &=\frac{1}{M_p}, \quad \text{if} \quad  r_c\: \&\: r_{c'}\geq \langle r \rangle \quad \text{or} \quad r_c\: \& \:r_{c'}< \langle r \rangle, \\
M_{cc'} &=0, \quad \text{otherwise}.
\end{aligned}
\end{equation}\\

For the example above, with four economies and six activities, $M_{cc'}$ takes the form:

\begin{equation}
M_{cc'}= \begin{bmatrix}
1/2 & 1/2 & 0 & 0   \\
1/2 & 1/2 & 0 & 0   \\
0 & 0 & 1/2 & 1/2  \\
0 & 0 & 1/2 & 1/2
\end{bmatrix}.
\end{equation}

Since we know the first eigenvector of this matrix is the vector $e^1_c=\mathbf{1}$, and since this matrix is symmetric, and has therefore orthogonal eigenvectors, we can use these properties to find the second eigenvector, which is:

\begin{equation}
e^2_{c}= ECI= \begin{bmatrix}
1    \\
1  \\
-1  \\
-1  
\end{bmatrix}.
\end{equation}\\

In this case, this eigenvector is also associated with the eigenvalue of one (this matrix is degenerate, meaning that it has more than one eigenvector associated with the same eigenvalue).\footnote{In this case, all linear combinations of these eigenvectors are eigenvectors themselves. For example the vector $[a,a,b,b]$ is also an eigenvector, since we can construct it as a linear combination of $[1,1,1,1]$ and $[1,1,-1,-1]$.} This eigenvector is easy to verify through multiplication.\\

What is important for us is that this eigenvector separates economies with above and below average $r$, that is:\\

\begin{equation}
\begin{aligned}
e^2_{c}&=ECI_c=1, \quad &\text{if} \quad r_c\geq\langle r \rangle,\\
e^2_{c}&=ECI_c=-1, \quad &\text{otherwise},
\end{aligned}
\end{equation}\\

showing that in this example the second eigenvector of the $M_{cc'}$ matrix or $ECI$ separate economies that are above or below average in their probability of having the only capability in the model.\footnote{At this point it is worth noting that a standard property of eigenvectors is that they have a freedom of sign. That is, if $e_c$ is an eigenvector of a matrix $M$ so is $-e_c$. This is trivial from the fact that if $Me_c=\lambda e_c$ then $M(-e_c)=\lambda (-e_c)$. This means that the eigenvector derivation of $ECI$ separates among economies based on their capability level, but is agnostic about which of the two clusters is the high-capability cluster. In the empirical literature, this is solved by iterating the system of eqns.~\eqref{eq:eci_system} to estimate $ECI$ starting from an initial condition that is correlated with the high-capability cluster (e.g. initializing the system with diversity $M_c$) and stopping at an even iteration. Other methods to estimate complexity empirically (e.g.~\cite{atkin2021globalization} also rely on an initialization guess).}\\

\begin{figure}[htb]
    \centering
    \includegraphics[width=0.7\textwidth]{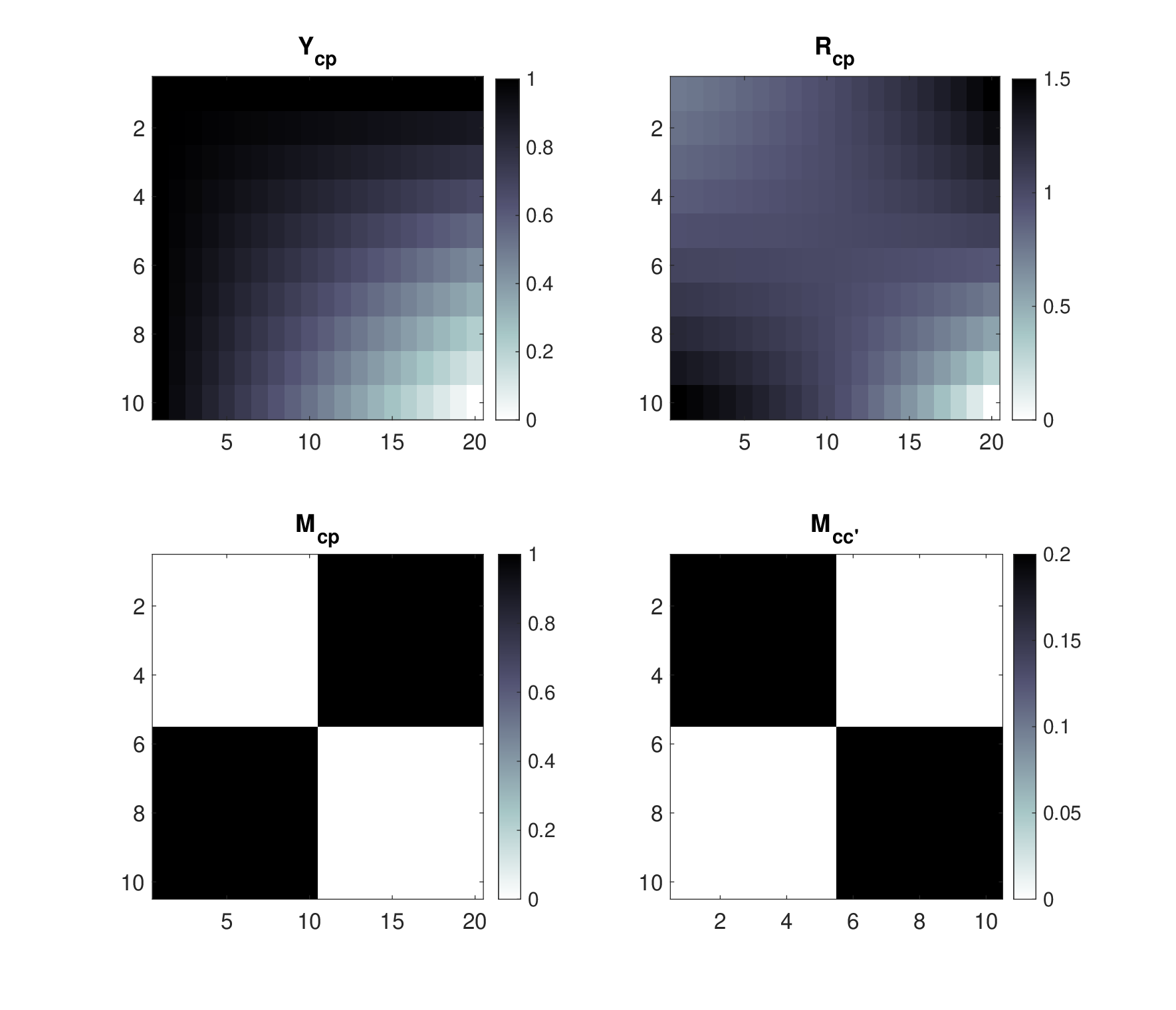}
    \caption{Graphical description of the four matrices involved in the single capability model for 10 countries and 20 products. In $cp$ matrices row represents economies (countries) and columns represent activities (products). Rows are sorted from highest $r_c$ to lowest $r_c$ and columns are sorted from lowest $q_p$ to highest $q_p$. That is, cell $(1,1)$ is the output of the country with the highest probability of having the capability on the product with the lowest probability of requiring it, and cell $(10,20)$ is the output of the country with the lowest probability of having a capability in the product with the highest probability of requiring it.}
    \label{fig:single_cap}
\end{figure}

Figure~\ref{fig:single_cap} visualizes the matrices in the single capability model for a case involving an even number of economies and activities (10 economies and 20 activities). These graphical representations will help us develop our intuition when interpreting more complex models later.\\

From top left to bottom right, we start with the output matrix ($Y_{cp}$), the specialization or RCA matrix ($R_{cp}$), the binary specialization matrix $M_{cp}$, and the complexity matrix $M_{cc'}$ from which we derive $ECI$. The output matrix $Y_{cp}$ shows a nested pattern, which is a tendency for the rows that are less filled to be subsets of the rows that are more filled. Nestedness is a well-known feature of matrices summarizing the geography of fine-grained economic activities, such as exports by country and product, employment by city and industry, or patents by city and technology~\cite{bustos_dynamics_2012}. It is also a common feature of bipartite networks in ecology (e.g. pollinator networks or geographic specialization networks\cite{mariani_nestedness_2019,almeida2008consistent}). This example shows how these transformations simplify $Y_{cp}$, reducing it to a couple of clusters with above and below average probability of having a capability. Yet, the symmetry of this example limits our ability to explore key properties of the method, such as the ability to separate capabilities from simple measures of diversity. For that, we need to consider other cases.\\

Next, we focus on the case where the number of economies is odd and the number of activities is even (and where the averages of $r$ and $q$ are still their medians). For example, $N_c=5$ and $N_p=6$. This example is interesting, because unlike in the previous case where the diversity of economies and the ubiquity of activities was constant, here only the ubiquity of activities remains fixed. This example is important because it will teach us about the ability of $ECI$ to recover $r_c$, even when the most diverse economy is the one that has a probability of having a capability equal to the average ($r_c=\langle r \rangle$) (it is actually specialized in all activities).\\

In this \emph{odd-even} case, $M_{cp}$ is given by:\\

\begin{equation}
\begin{aligned}
M_{cp} &= 1,\quad \text{if} \quad  r_c\: > \langle r \rangle \quad \& \quad q_p > \langle q \rangle, \\
M_{cp} &= 1,\quad \text{if} \quad  r_c\: < \langle r \rangle \quad \& \quad q_p < \langle q \rangle, \\
M_{cp} &= 1,\quad \text{if} \quad  r_c\: = \langle r \rangle \quad, \\
M_{cp} &=0, \quad \text{otherwise},
\end{aligned}
\end{equation}\\

which for $N_c=5$ and $N_p=6$ results in the binary specialization matrix that is completely filled on the third row (so the matrix is no longer symmetric):
\begin{equation}
M_{cp}= \begin{bmatrix}
0 &0 & 0 & 1 & 1 & 1  \\
0 &0 & 0 & 1 & 1 & 1  \\
1 & 1 & 1 & 1 & 1 & 1  \\
1 & 1 & 1 & 0 &0 & 0   \\
1 & 1 & 1 & 0 &0 & 0 
\end{bmatrix}.
\end{equation}\\
Clearly the most diverse economy is the one in the third row, which is specialized in all activities.\\

\begin{figure}[htb]
    \centering
    \includegraphics[width=0.7\textwidth]{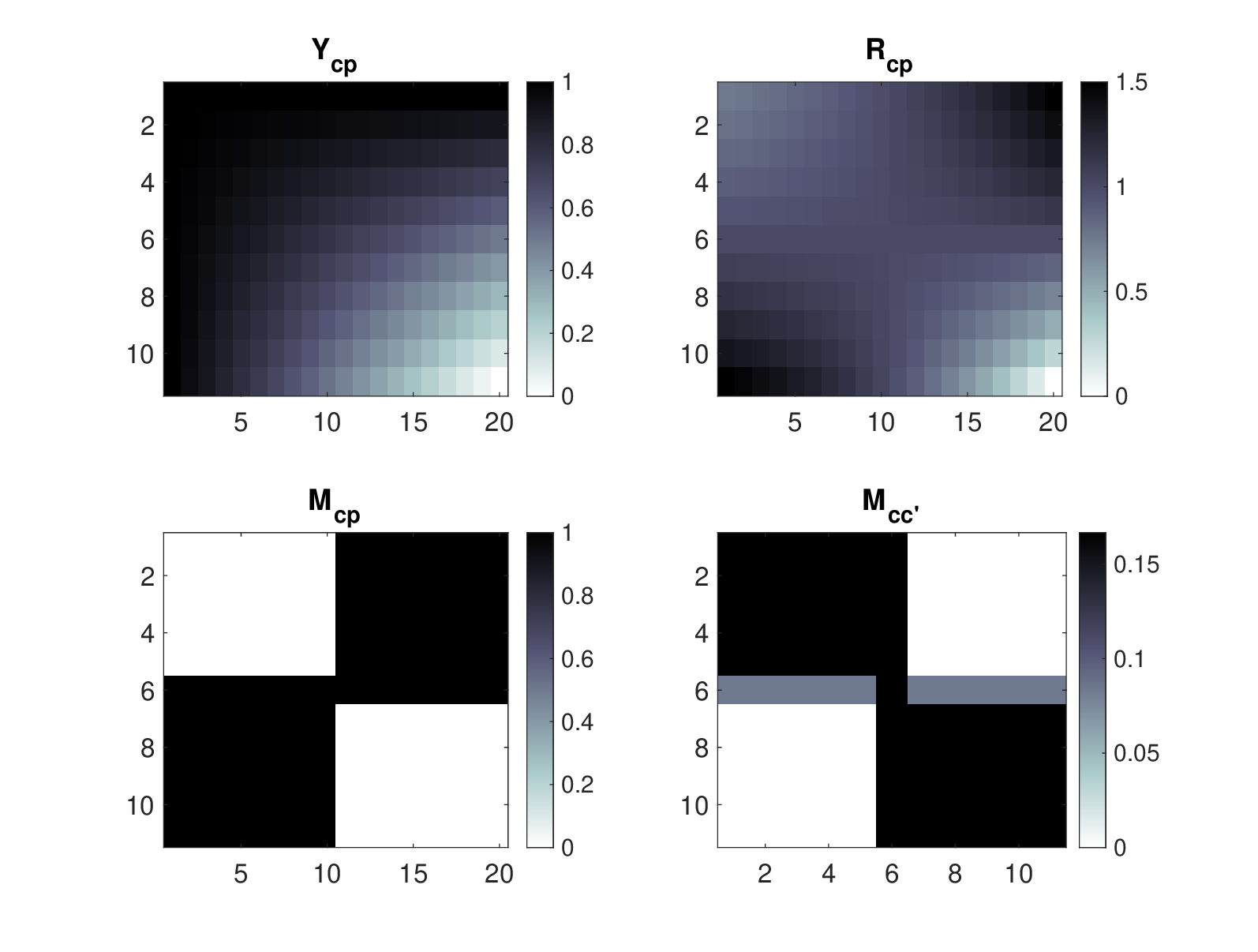}
    \caption{Graphical description of the four matrices involved in the single capability model for 11 economies (e.g. countries) and 20 activities (e.g products). In $cp$ matrices row represents economies (countries) and columns represent activities (products). Rows are sorted from highest $r_c$ to lowest $r_c$ and columns are sorted from lowest $q_p$ to highest $q_p$. That is, cell $(1,1)$ is the output of the country with the highest probability of having the capability on the product with the lowest probability of requiring it, and cell $(11,20)$ is the output of the country with the lowest probability of having a capability in the product with the highest probability of requiring it.}
    \label{fig:single_cap2}
\end{figure}

Moving to $M_{cc'}$ gives us:\\
\begin{equation}
\begin{aligned}
M_{cc'} &=\frac{1}{M_p}, \quad \text{if} \quad c=c',\\
M_{cc'} &=\frac{1}{M_p}, \quad \text{if} \quad  r_c\: \&\: r_{c'}> \langle r \rangle \quad \text{or} \quad r_c\: \& \:r_{c'}< \langle r \rangle, \\
M_{cc'} &=\frac{1}{M_c}\sum_p\frac{{M_{cp}M_{c'p}}}{M_p}, \quad \text{if} \quad  r_c\: \&\: r_{c'} = \langle r \rangle \quad \& \quad c \neq c',\\
M_{cc'} &=0, \quad \text{otherwise},
\end{aligned}  
\end{equation}\\

which for five economies and six activities results in the matrix:
\begin{equation}
M_{cc'}= \begin{bmatrix}
1/3 & 1/3 & 1/3 & 0 & 0  \\
1/3 & 1/3 & 1/3 & 0 & 0  \\
1/6 & 1/6 & 1/3 & 1/6 & 1/6  \\
0 & 0 & 1/3 & 1/3 & 1/3  \\
0 & 0 & 1/3 & 1/3  & 1/3
\end{bmatrix}.
\end{equation}\\
This matrix is also quite regular, and has the following second eigenvector which can be verified simply using matrix multiplication:
\begin{equation}
e^2_{c}= ECI_c=\begin{bmatrix}
1    \\
1  \\
0\\
-1  \\
-1  
\end{bmatrix}.
\end{equation}\\
In  more general terms it is given by:\\
\begin{equation}
\begin{aligned}
e^2_c &= ECI_c = 1,  \quad &\text{if} \quad  r_c > \langle r \rangle, \\
e^2_c &= ECI_c = -1, \quad &\text{if} \quad r_c< \langle r \rangle, \\
e^2_c &= ECI_c = 0, \quad &\text{if} \quad r_c = \langle r \rangle.
\end{aligned}    
\end{equation}\\

This is an interesting result, since it shows that the second eigenvector or $ECI$ is not ``fooled by diversity.'' On the contrary, it is able to recover the fact that the economy that is specialized in all activities has a probability of having a capability that is in between that of the high probability and low probability clusters.\\ 

Figure~\ref{fig:single_cap2} summarizes the matrices in the single capability model for a case involving an odd number of economies and an even number of activities (11 economies and 20 activities). In this case, the key difference is that center row of $M_{cp}$ which extends through all columns of the matrix and results in a small overlap between the two clusters in $M_{cc'}$.\\

Finally, we consider the case in which both the number of economies and activities are odd, which, as we will see, points towards the robustness when both the economy and activity distributions are asymmetric, introducing an activity with average factor intensity alongside the economy with average capability level. In this case, the diversity of economies and the ubiquity of activities is no longer constant. Now the $M_{cp}$ matrix has both, one row and one column that are completely filled, which correspond respectively to the economy and activity with $r_{c} = \langle r \rangle$ and $q_{p} = \langle q \rangle$. That is:\\

\begin{equation}
\begin{aligned}
M_{cp} &= 1,\quad \text{if} \quad  r_c> \langle r \rangle \quad \& \quad q_p\: > \langle q \rangle, \\
M_{cp} &= 1,\quad \text{if} \quad  r_c< \langle r \rangle \quad \& \quad q_p\:  < \langle q \rangle, \\
M_{cp} &= 1,\quad \text{if} \quad  r_c\: = \langle r \rangle, \\
M_{cp} &= 1,\quad \text{if} \quad  q_p\: = \langle q \rangle,\\
M_{cp} &=0, \quad \text{otherwise},
\end{aligned}
\end{equation}\\

which we can bring to an example with five economies and seven activities:\\

\begin{equation}
M_{cp}= \begin{bmatrix}
0 &0 & 0 & 1 & 1 & 1 & 1  \\
0 &0 & 0 & 1 & 1 & 1 & 1 \\
1 & 1 & 1 & 1 & 1 & 1 & 1 \\
1 & 1 & 1 & 1 &0 & 0  & 0 \\
1 & 1 & 1 & 1 &0 & 0  & 0
\end{bmatrix}.
\end{equation}\\

In this case $M_{cc'}$ will have a more complex form which we can express by noticing that the diversity and ubiquity of the economy and activity in the middle row and column of $M_{cp}$ is the number of economies $N_c$ and the number of activities $N_p$. Since all other economies and activities have the same diversity and ubiquity, which we will denote by $M_c$ and $M_p$, we obtain:\\

\begin{equation}
\begin{aligned}
M_{cc'} &=\frac{1}{M_c}\left(1+\frac{1}{N_p}\right), \quad \text{if}\quad  r_c\: \&\: r_{c'}> \langle r \rangle \quad \textrm{or} \quad r_c\: \& \:r_{c'}< \langle r \rangle,\\
M_{cc'} &=\frac{1}{M_c N_p}, \quad \text{if}\quad  r_c\: > \langle r \rangle \quad \& \quad r_{c'}< \langle r \rangle\ \text{and vice versa},\\
M_{cc'} &=\frac{1}{N_c}\left(\frac{1}{N_p}+\frac{N_c-1}{M_p}\right), \quad \text{if}\quad  c\: =\:c' \quad  \& \quad  r_c\: = \langle r \rangle,\\
M_{cc'} &=\frac{1}{N_c}\left(1+\frac{1}{N_p}\right), \quad \text{if}\quad  c\: \neq \:c'\quad \& \quad  r_c\: = \langle r \rangle,\\
M_{cc'} &=\frac{1}{M_c}\left(1+\frac{1}{N_p}\right), \quad \text{if}\quad  r_{c'}\: = \langle r \rangle,
\end{aligned}
\end{equation}
\\

which might be easier to parse when presented in matrix form:\\
\begin{equation}
\renewcommand{\arraystretch}{1.7}
M_{cc'}= \begin{bmatrix}
\frac{1}{M_c}\left(\frac{N_p+1}{N_p}\right) &  \dots & \frac{1}{M_c}\left(\frac{N_p+1}{N_p}\right)  & \dots & \frac{1}{M_c N_p}  \\

\frac{1}{M_c}\left(\frac{N_p+1}{N_p}\right) &  \dots & \frac{1}{M_c}\left(\frac{N_p+1}{N_p}\right)  & \dots & \frac{1}{M_c N_p}  \\

 \frac{1}{N_c}\left(1+\frac{1}{N_p}\right) & \dots & \frac{1}{N_c}\left(\frac{1}{N_p}+\frac{N_c-1}{M_p}\right) & \dots & \frac{1}{N_c}\left(1+\frac{1}{N_p}\right)  \\

 \frac{1}{M_c}\left(\frac{1}{N_p}\right) & \dots & \frac{1}{M_c}\left(\frac{N_p+1}{N_p}\right)  & \dots & \frac{1}{M_c}\left(\frac{N_p+1}{N_p}\right)   \\

 \frac{1}{M_c}(\frac{1}{N_p}) & \dots & \frac{1}{M_c}(\frac{N_p+1}{N_p}) &\dots & \frac{1}{M_c}(\frac{N_p+1}{N_p}) 
\end{bmatrix}.
\renewcommand{\arraystretch}{1}
\label{oddoddmatrix}
\end{equation}\\
Bringing this to the five economies and seven activities example gives us:\\
\begin{equation}
M_{cc'}= \begin{bmatrix}
3/10 & 3/10 & 3/10 & 1/20 & 1/20  \\
3/10 & 3/10 & 3/10 & 1/20 & 1/20  \\
6/35 & 6/35 & 11/35 & 6/35 & 6/35 \\
1/20 & 1/20 & 3/10 & 3/10 & 3/10   \\
1/20 & 1/20 & 3/10 & 3/10 & 3/10  
\end{bmatrix},
\end{equation}\\
which again has a second eigenvector of the form:
\begin{equation}
e^2_{c}= ECI_c =\begin{bmatrix}
a    \\
a  \\
0\\
-a  \\
-a  
\end{bmatrix}.
\end{equation}\\

\begin{figure}[htb]
    \centering
    \includegraphics[width=0.7\textwidth]{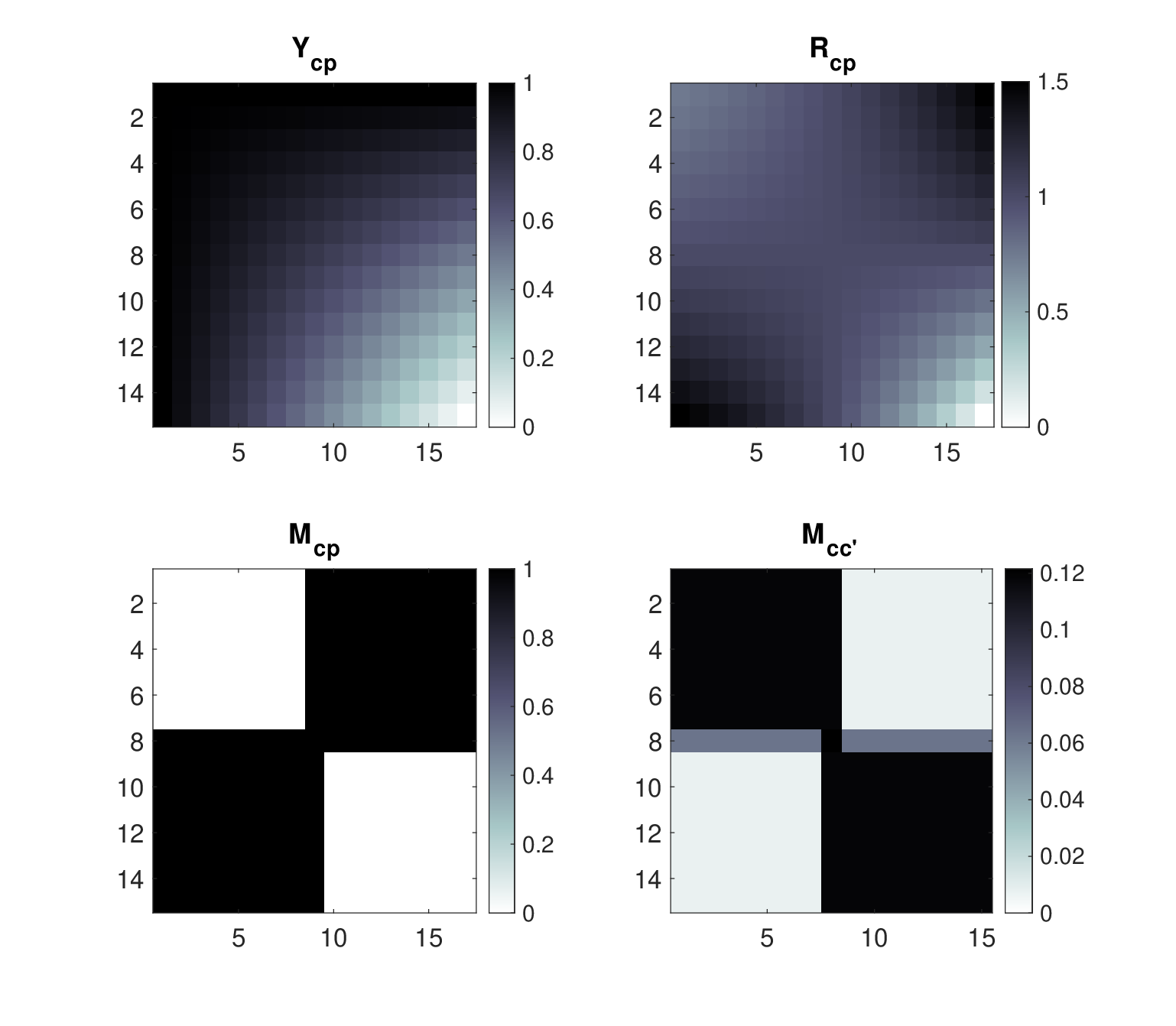}
    \caption{Graphical description of the four matrices involved in the single capability model for 15 economies (e.g. countries) and 17 activities (e.g products). In $cp$ matrices row represents economies (countries) and columns represent activities (products). Rows are sorted from highest $r_c$ to lowest $r_c$ and columns are sorted from lowest $q_p$ to highest $q_p$. That is, cell $(1,1)$ is the output of the economy or country with the highest probability of having the capability on the product or activity with the lowest probability of requiring it, and cell $(11,20)$ is the output of the economy with the lowest probability of having a capability in the activity or product with the highest probability of requiring it.}
    \label{fig:single_cap3}
\end{figure}

This is easy to verify through multiplication. Since the vector adds all of the elements up to the center column and then subtracts all of the elements after the central column, and since the number of elements before and after the central column are the same, we can simply subtract the first and last element of the first row of matrix (eqn. \ref{oddoddmatrix}) to obtain:\\
\begin{equation}
M_{cp}v_c=\frac{a}{M_c}\left(1+\frac{1}{M_p}\right)-\frac{a}{M_c}\frac{1}{M_p}=\frac{a}{M_c}.
\end{equation}\\
Doing the same operation on the last row we get:\\
\begin{equation}
M_{cp}v_c=\frac{a}{M_c}\left(\frac{1}{M_p}\right)-\frac{a}{M_c}\left(1+\frac{1}{M_p}\right)=-\frac{a}{M_c}.
\end{equation}\\
Since in the central row of the matrix all elements, except the one in the diagonal, are the same, this vector sends that row to zero. Thus, up to a normalization constant, the second eigenvector of $M_{cc'}$ is given by:\\
\begin{equation}
\begin{aligned}
e^2_c &= ECI_c = a,  \quad &\text{if} \quad  r_c> \langle r \rangle,  \\
e^2_c &= ECI_c = -a, \quad &\text{if} \quad r_c< \langle r \rangle, \\
e^2_c &= ECI_c = 0, \quad &\text{if} \quad r_c= \langle r \rangle.
\end{aligned}
\end{equation}\\

Figure~\ref{fig:single_cap3} presents these matrices in graphical form. We would like to notice two things about this version of the single capability model. The first one is that in this case $M_{cc'}$ no longer has blocks of $0$s. The second one is that this is also an example in which the highest diversity economy (the $3^{rd}$ row in $M_{cp}$ is correctly identified as not being the economy with the highest probability of having the capability.\\ 

Thus, we have shown that, in the context of the single capability or single factor model, the second eigenvector of the $M_{cc'}$ matrix, known as the economic complexity index or $ECI$, separates economies among those that have a higher and lower than average probability of having the single capability in the model.\\

In the next section we will use similar figures to explore more complex forms of this model, involving multiple and heterogeneous capabilities. We will then move to different production functions to explore the generalizability of this result.\\

\section{The Multi Capability Model}
\label{sec:multi_capability-model}

The multi capability version of the combinatorial model can be defined by letting the probability that a country has capability $b$ be $r_{c,b}$ and the probability that a product requires a capability $b$ be $q_{p,b}$. For a country to produce a product it needs to have all of the capabilities that the product requires. That is, the product of these probabilities for all of the capabilities in the model. Mathematically, that translates into an output matrix of the form:\footnote{This model assumes capabilities are not substitutable. A model with substitutable capabilities would take the form \begin{equation}
    Y_{cp}=A_{cp}\prod_{b=1}^{N_b}(1-q_{p,b}(1-r_{c,b}-\sum_{b'\neq b}S_{bb'}r_{cb'}))
\end{equation}
where $S_{bb'}$ is a matrix describing the level of substitutability between capabilities $b$ and $b'$.}\\

\begin{equation}
    Y_{cp}=A\prod_{b=1}^{N_b}(1-q_{p,b}(1-r_{c,b})).
    \label{multi_capability_model}
\end{equation}\\

To avoid over-parameterizing the model too early, and to simplify our exploration, we will begin with the case in which these probabilities are independent of the capability index and of each other, and where the pre-factor $A_{cp}$ is constant. That is:\\

\begin{equation}
    Y_{cp}=A\prod_{b=1}^{N_b}(1-q_{p}(1-r_{c})),
    \label{multi_capability_model_simple}
\end{equation}\\
which reduces to a well-known binomial form\footnote{While this form looks relatively simple, even the solution for $N_b=2$ can result in a mathematical form that is substantially more complicated than the one for the single-capability model. In fact, after some algebra one can show that the condition for $R_{cp}\geq1$ in the $N_b=2$ case is:
\begin{equation}
\begin{aligned}
\left[q_p^2-\langle q^2 \rangle\right]\left[r_c^2-\langle r^2 \rangle + 2(r_c-\langle r \rangle)\right]+2\left[q_p - \langle q \rangle\right]\left[r_c - \langle r\rangle\right]+\\
2\left[q_p\langle q^2 \rangle - \langle q\rangle q_p^2\right]\left[r_c-\langle r \rangle r_c^2\langle r \rangle - \langle r^2\rangle r_c +\langle r^2 \rangle - r_c^2\right] \geq 0.
\end{aligned}
\end{equation}
}:\\
\begin{equation}
    Y_{cp}=(1-q_{p}(1-r_{c}))^{N_b}
    \label{multi_capability_model_exp}
\end{equation}\\

This form assumes that a country has the same probability of having each of the different capabilities required by a product. The need for multiple capabilities, therefore, enters only in the probability of missing one of them, making this similar in spirit to Kremer's O-Ring model~\cite{kremer_o-ring_1993}. In fact, Kremer's O-Ring production function can be recovered from eqn.(\ref{multi_capability_model_simple}) by setting $q_{p,b}=1$ for all activities and $r_{c,b}=r_c$ for each economy.\footnote{In that case, the production function reduces to the Kremer-Shockley function:
\begin{equation}
Y_c=A\prod_{b=1}^{N_b}r_c=r_c^{N_b}.
\end{equation}}\\ 

\begin{figure}[htbp]
    \centering
    \includegraphics[width=0.8\textwidth]{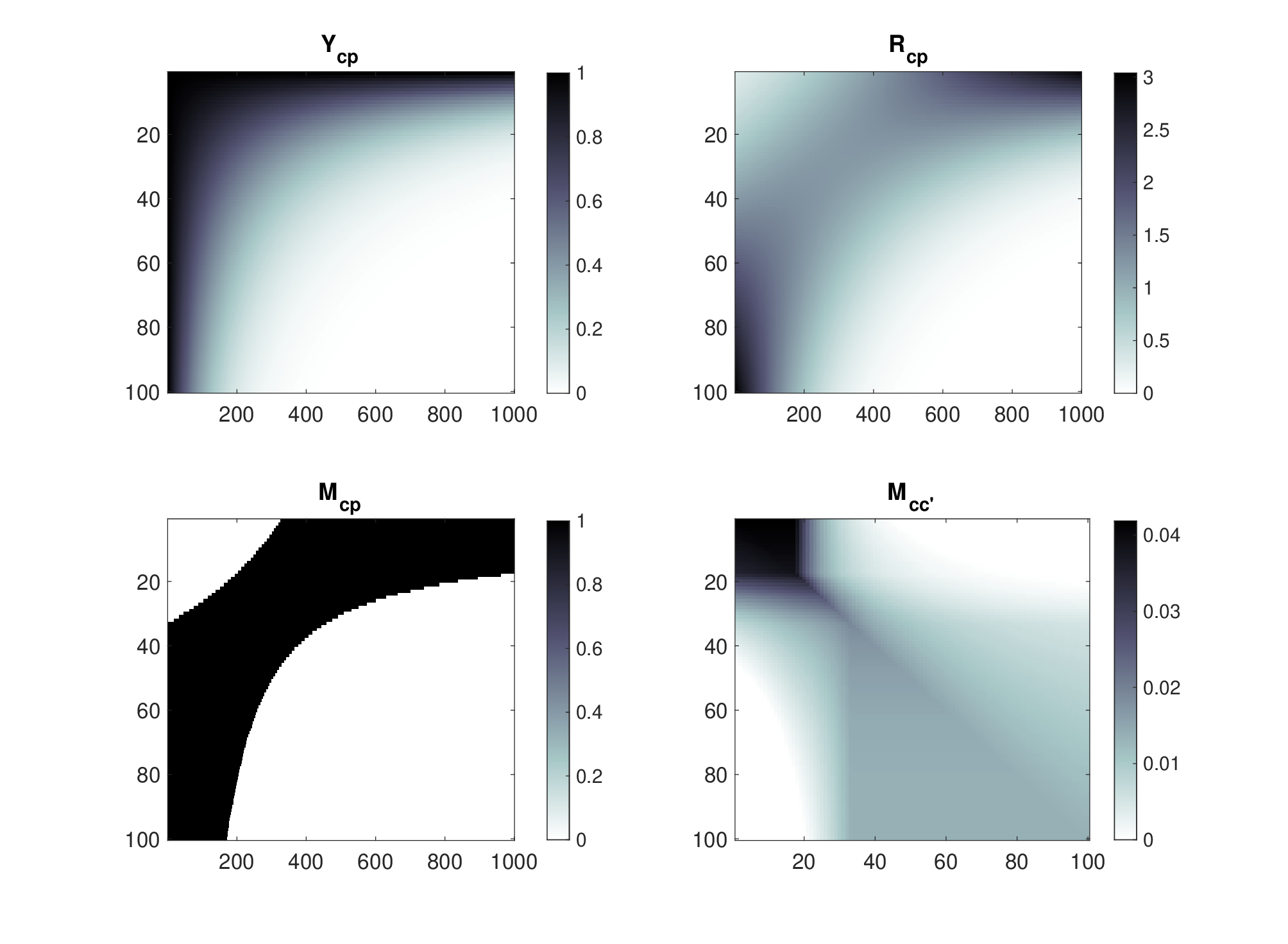}
    \caption{The four matrices involved in economic complexity calculations using a ten capability model for 100 countries and 1,000 products. In $cp$ matrices row represents economies (countries) and columns represent activities (products). Rows are sorted from highest $r_c$ to lowest $r_c$ and columns are sorted from lowest $q_p$ to highest $q_p$.}
    \label{fig:multi_cap50}
\end{figure}

Figure~\ref{fig:multi_cap50} shows the same matrices we derived analytically for the single capability model but for a model involving ten capabilities, one hundred economies, and one thousand activities. This gets to a granularity that is similar to the one used in empirical economic complexity studies.\\

In this example, economies and activities are modeled using evenly spaced probabilities in the $[0,1]$ interval. That is, for an eleven economy model the probabilities are given by ${0,0.1,0.2,\dots,0.9,1}$. The result is a highly nested output matrix $Y_{cp}$ and strongly off-diagonal specialization matrices ($R_{cp}$ and $M_{cp}$).\\

It is worth noting that the more diverse economies in this model are not the ones with the highest $r_c$, but the ones with an $r_c$ that is below the largest (around $0.8$). This is because the reduced output of these economies in the most demanding activities (the ones with highest $q_p$) means they are relatively more specialized in products with lower $q_p$s compared to the highest $r_c$ economies. This effect is analogous to what we saw in the one capability model when we considered an odd number of economies.\\

Figure~\ref{fig:multi_cap50} also shows that $M_{cc'}$ follows a similar block diagonal structure than before, but much smoother than in the single capability model.\\

\begin{figure}[htbp]
    \centering
    \includegraphics[width=0.8\textwidth]{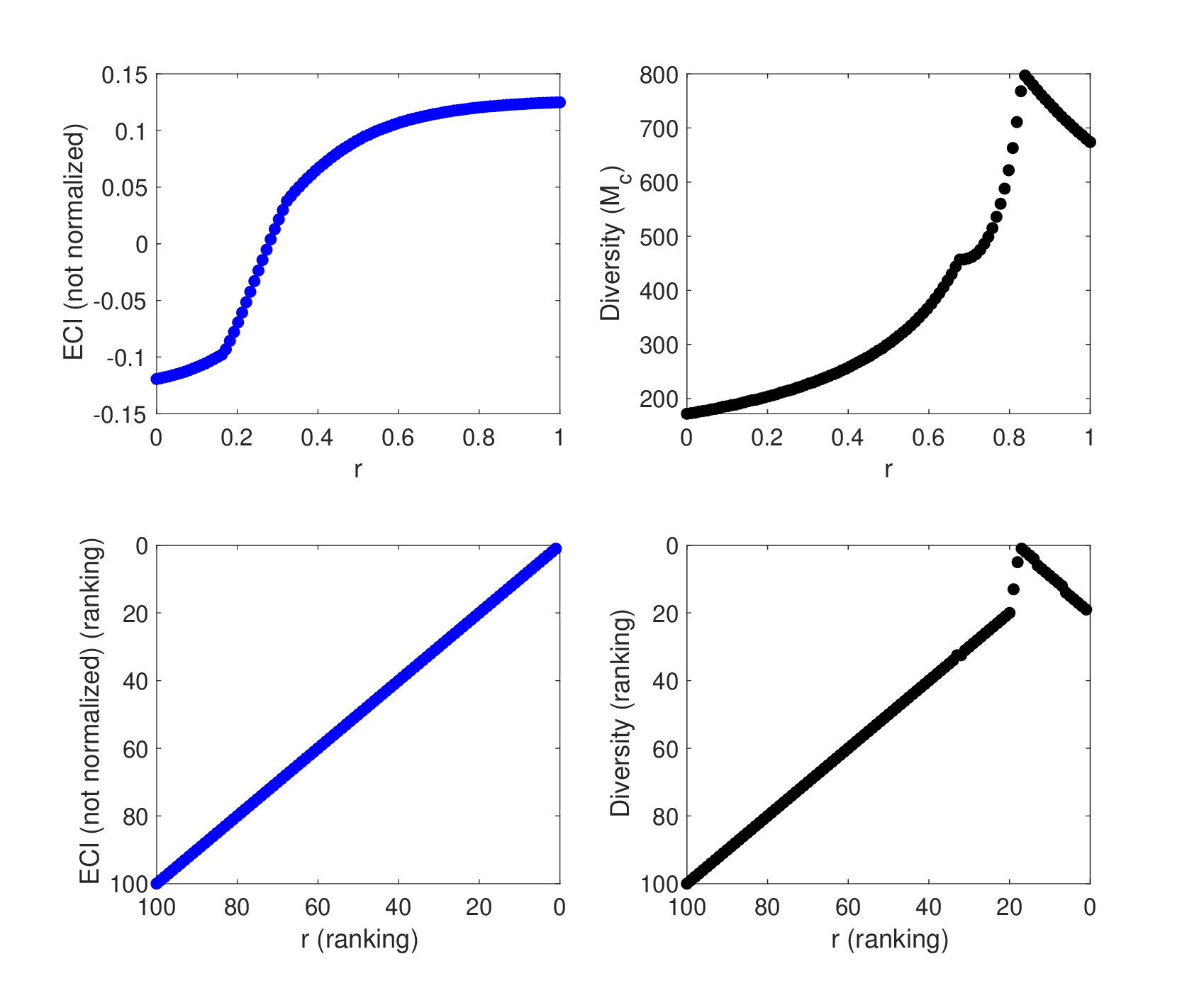}
    \caption{Comparison between the key parameters representing economies in the model ($r$), the second eigenvector of the $M_{cc'}$ matrix ($ECI$), and the diversity ($M_c$) of economies in the model. The top two panels show the raw relationship between the variables while the bottom two compare their rankings.}
    \label{fig:eigen_multi_cap50}
\end{figure}

While it would certainly be substantially more difficult to estimate the eigenvectors of this model analytically, we can still explore them numerically. Figure~\ref{fig:eigen_multi_cap50} compares the $r_c$ of each economy with its second eigenvector of the $M_{cc'}$ matrix (the non-normalized $ECI$), diversity ($M_{c}$), and the ranking of economies according to $ECI$. Unlike in the single capability model, where $ECI$ told us only if an economy was above or below average, in this example we get a less discrete second eigenvector that increases monotonically with $r$. This results in a perfect correlation between the ranked values of $r$ and $ECI$. Diversity, however, peaks for economies with $r_c$ less than the maximum, meaning that it is a non-ideal way to estimate the capabilities of economies in this model. That is, we recover the fact that the second eigenvector of $M_{cc'}$--the economic complexity index ($ECI$)--is a good method to estimate the key parameter for the economies in the model ($r_c$), extending our previous result to a multi-capability setting.

\begin{figure}[htbp]
    \centering
    \includegraphics[width=1.1\textwidth]{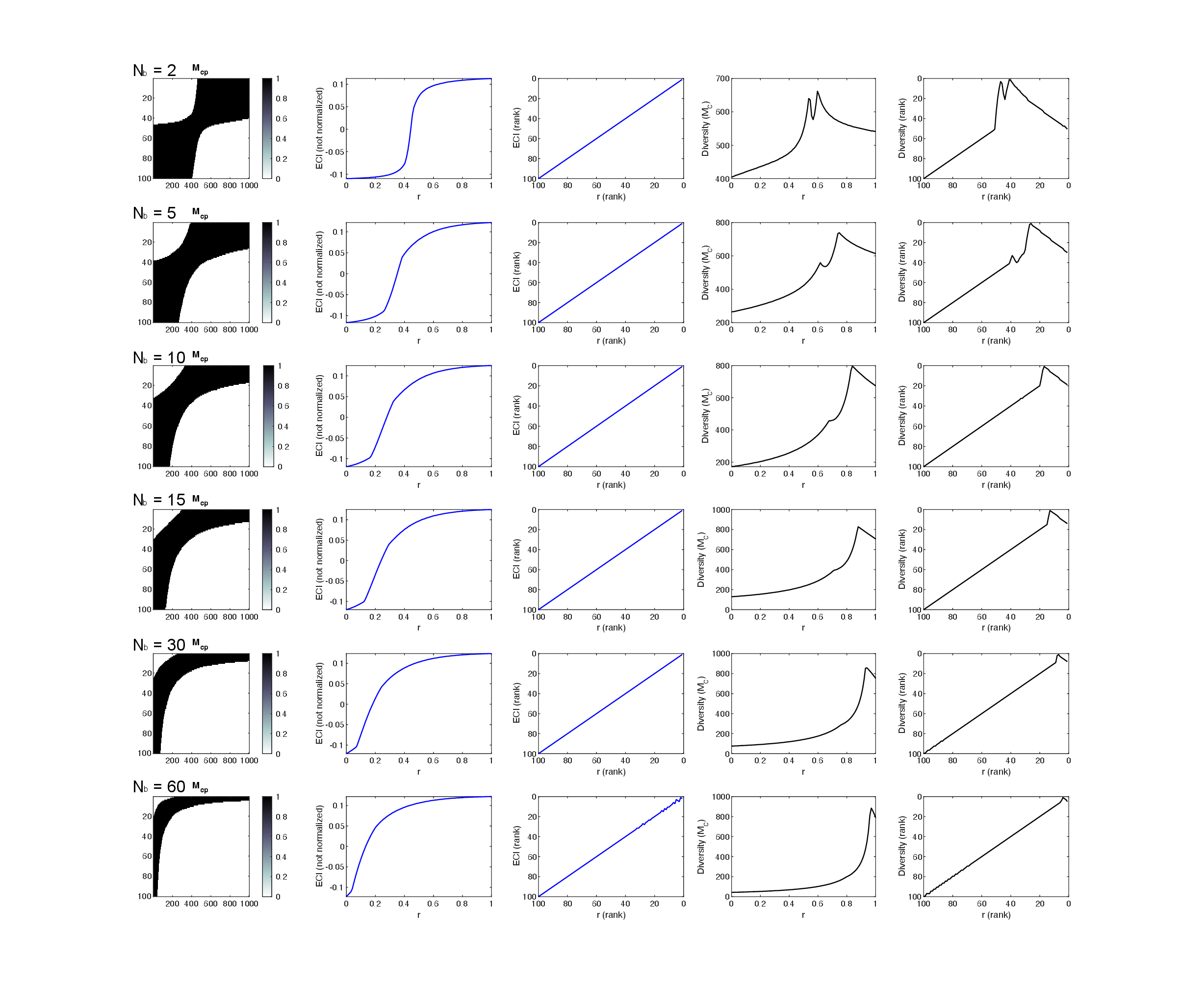}
    \caption{Comparison for the binary specialization matrix and the correlation between complexity, diversity, and the probability that a country has a capability $r$ for models using $2,5,10,15,30,$ and $60$ capabilities.}
    \label{fig:capabilitygrid}
\end{figure}

Figure~\ref{fig:capabilitygrid} illustrates the behavior of this model for different number of capabilities (from 2 to 60). Overall, the behavior observed is consistent with the one observed for the ten capability example. Across the board, $ECI$ behaves as a perfect estimator of the probability that a country has a capability. We can observe, however, that diversity improves as an indicator for the models with the highest number of capabilities (60), becoming almost perfectly monotonic in that case.\\

To continue our exploration we relax our assumptions about the distributions of $r_c$ and $q_p$. So far, our simulations have involved evenly spaced $r_c$s and $q_p$s in the $[0,1]$ interval, which are an idealized uniform distribution. So we replace these uniform distributions for Gaussians by drawing a random number from a normal distribution for each $r_c$ and $q_p$ and min-max normalizing these random numbers to ensure they fall in the $[0,1]$ interval.\\
\begin{figure}[htbp]
    \centering
    \includegraphics[width=0.8\textwidth]{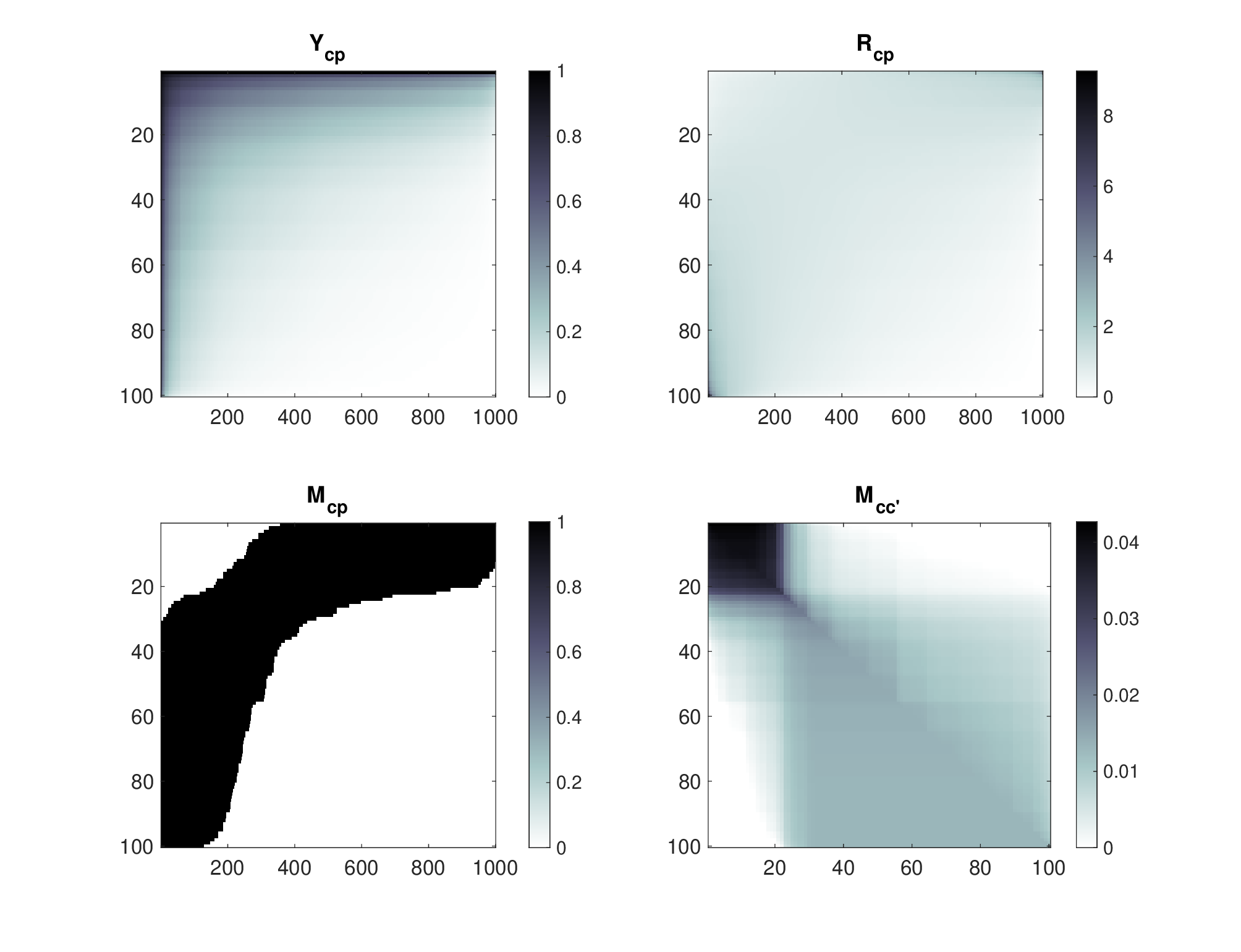}
    \caption{The four matrices involved in economic complexity calculations using a ten capability model for 100 economies and 1000 activities using randomly assigned $r_c$s and $q_p$s according to a normal distribution. In $cp$ matrices row represents economies (countries) and columns represent activities (products). Rows are sorted from highest $r_c$ to lowest $r_c$ and columns are sorted from lowest $q_p$ to highest $q_p$.}
    \label{fig:multi_capnorm}
\end{figure}

\begin{figure}[htbp]
    \centering
    \includegraphics[width=0.8\textwidth]{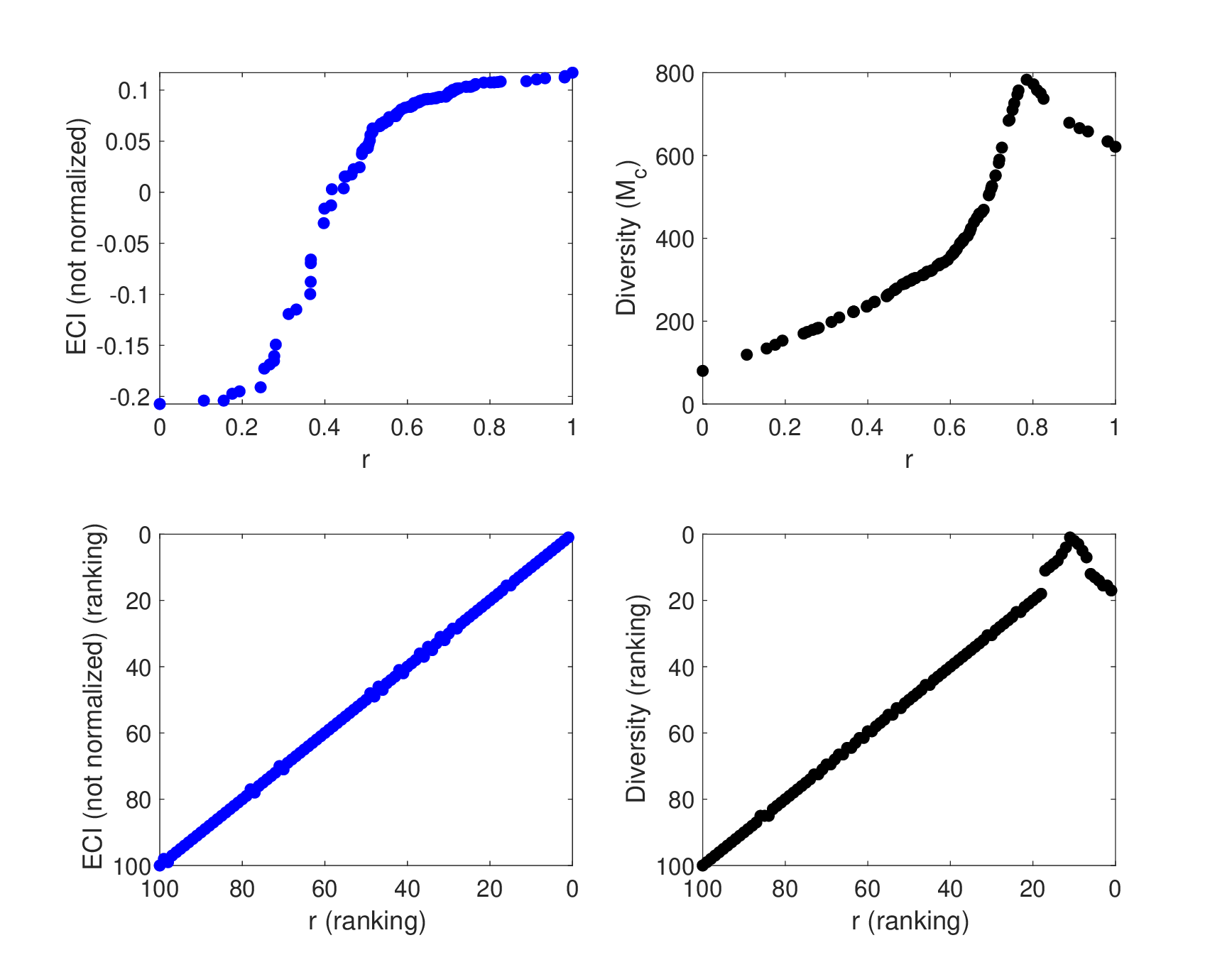}
    \caption{Comparison between the key parameter representing economies in the model ($r$), the second eigenvector of the $M_{cc'}$ matrix ($ECI$), and the diversity ($M_c$) of economies in a model involving 100 economies, 1000 activities, and 10 capabilities. The top two figures show the relationship between the raw variables and the bottom two show that relationship in rankings.}
    \label{fig:eigen_multi_norm}
\end{figure}

Figures~\ref{fig:multi_capnorm} and~\ref{fig:eigen_multi_norm} show the results of this exercise. Unlike in the previous example, the specialization matrix $M_{cp}$ exhibits a bit more ``roughness'', with non-perfectly smooth edges. That said, the behavior of this model is otherwise similar to the previous one. $M_{cc'}$ is roughly block diagonal and the second eigenvector of $M_{cc'}$ or $ECI$ almost perfectly captures $r_c$, as seen in its monotonic relationship with $r$ and in their rank correlation (Figure~\ref{fig:eigen_multi_norm}), whereas diversity peaks for economies with an $r_c$ of around $3/4$, making it a non-ideal estimator of $r_c$.\\

Now we consider the case in which the probability that an economy has a capability, and that an activity requires one, is not equal across all capabilities. That is, we consider heterogeneous capabilities of the form:

\begin{align}
    r_c \xrightarrow{} r_{c,b},\\ 
    q_p \xrightarrow{} q_{p,b}.
\end{align}

Using the following parametrization:

\begin{equation}
\label{eq:parametrization}
\begin{cases}
   r_{c,b} = \alpha r_c+(1-\alpha)\,\text{random}(0,1), \\[6pt]
   q_{p,b} = \alpha q_p+(1-\alpha)\,\text{random}(0,1).
\end{cases}
\end{equation}\\

That is, we set a baseline level for the probability that an economy has a capability or that an activity requires one (which we do through a linear function), and mix that with a random number according to the proportions $\alpha$ and $1-\alpha$. When $\alpha=1$ the probability that an economy has a capability is the same for all economies and we recover our previous case. When $\alpha=0$ the capability levels are fully random.\\

In this case, our goal is to explore whether $ECI$ is able to recover the underlying structure of capabilities. So, we compare $ECI$ with both the average level of capabilities $\langle r_c\rangle = \sum_b r_{cb} / N_b$ and the leading singular vector of the capability matrix $r_{cb}$. The average captures the overall level of capabilities in each country, whereas the leading singular vector identifies the dominant mode of variation--the main direction along which countries differ in their capability profiles.\\

\begin{figure}[tbp]
    \centering
    \includegraphics[width=0.8\textwidth]{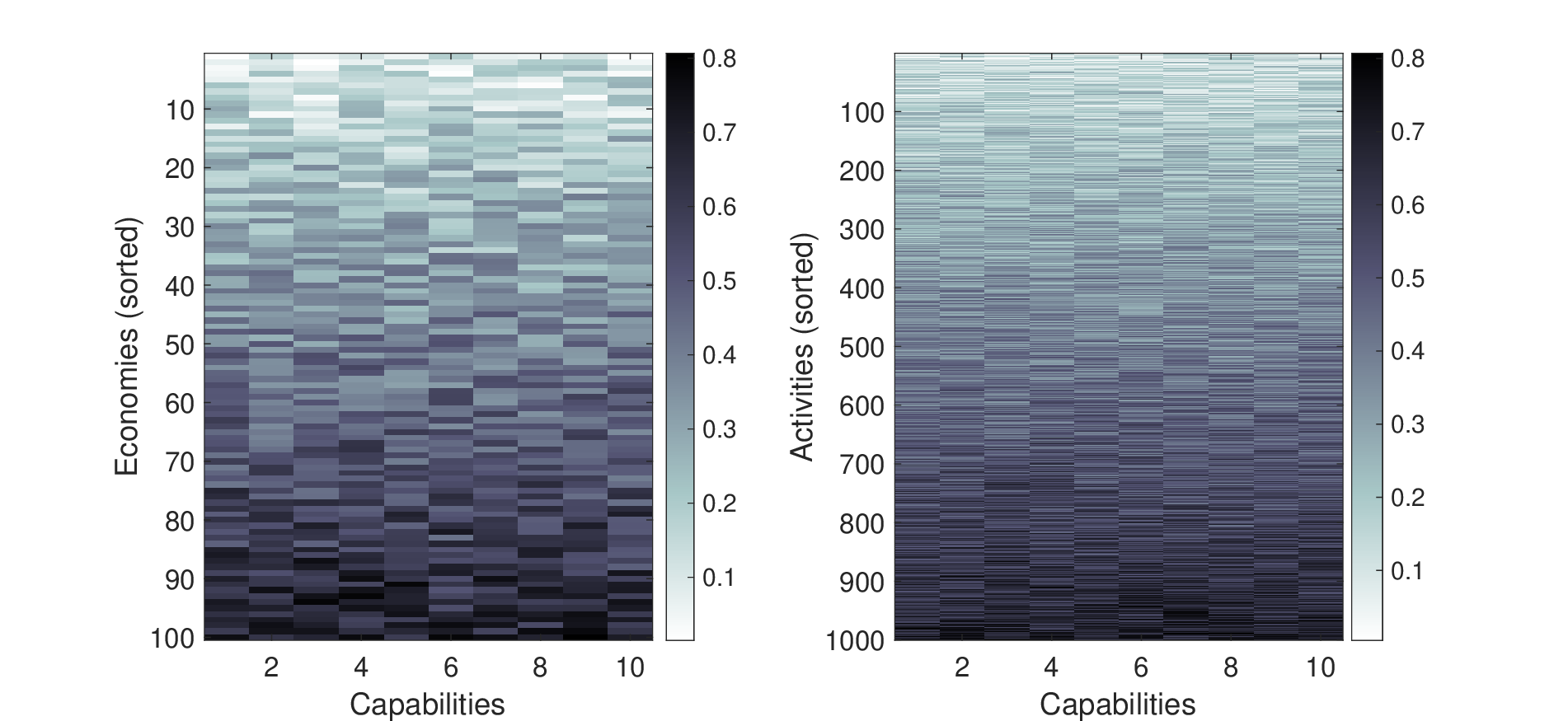}
    \caption{Parametrization of $r_{c,b}$ and $q_{p,b}$ for ten capabilities in a model where the probability that an economy has a capability, or that an activity requires it, is $3/4$ of a linearly spaced baseline in the $[0,1]$ interval and $1/4$ random.}
    \label{parametrization}
\end{figure}

\begin{figure}[htbp]
    \centering
    \includegraphics[width=0.8\textwidth]{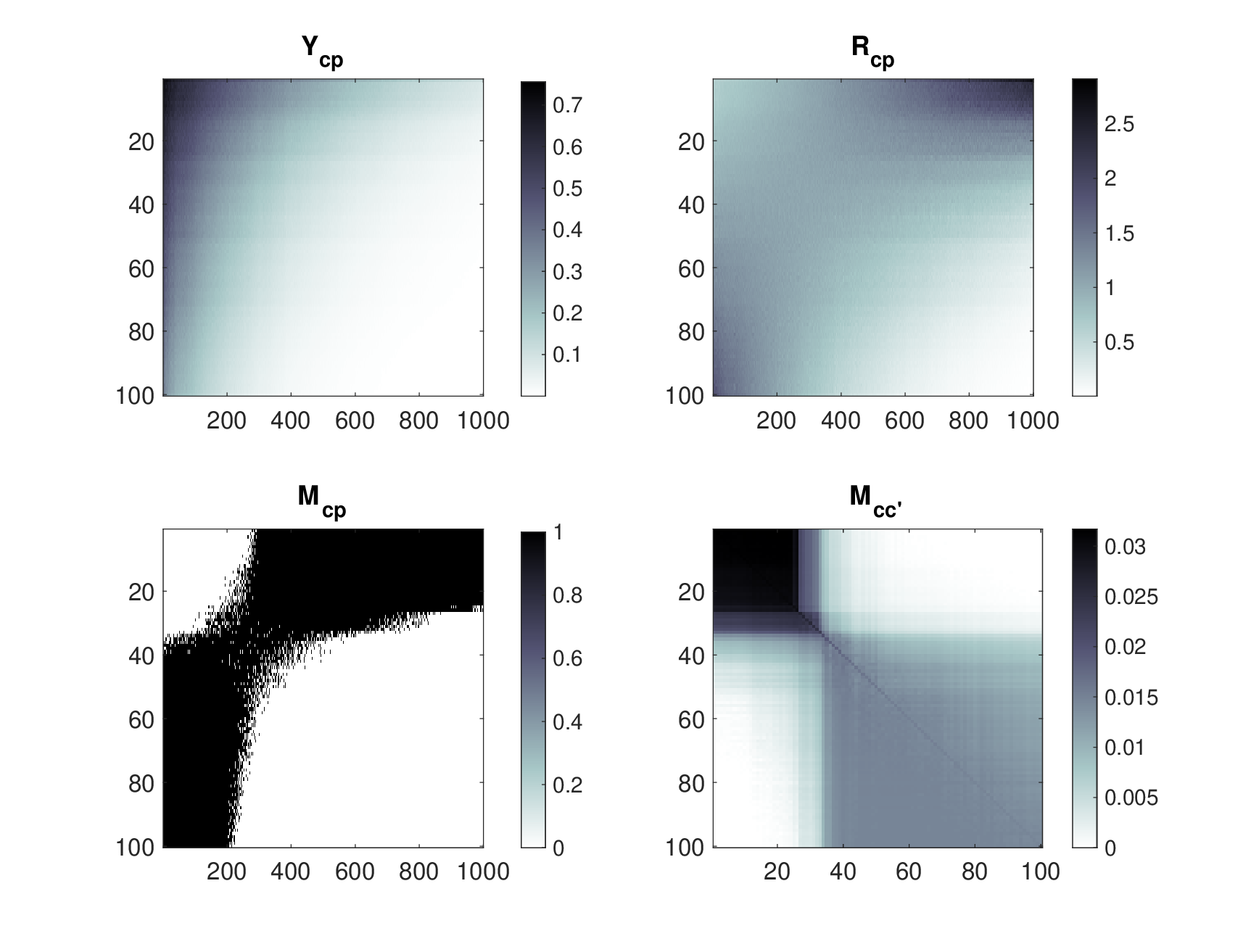}
    \caption{Matrices for a 10 capability, 100 economies, and 1000 activities model, where the probability that an economy has a capability, or that an activity requires it, is $3/4$ of a linearly spaced baseline in the $[0,1]$ interval and $1/4$ random.}
    \label{matricesdiffcap}
\end{figure}

Figure~\ref{parametrization} provides an illustration of this parametrization for the case when the probability that an economy has a capability, or that an activity requires it, is $3/4$ of a linearly spaced baseline in the $[0,1]$ interval and $1/4$ random. The matrices resulting from this model are shown in Figure~\ref{matricesdiffcap}.\\

\begin{figure}[htbp]
    \centering
    \includegraphics[width=0.99\textwidth]{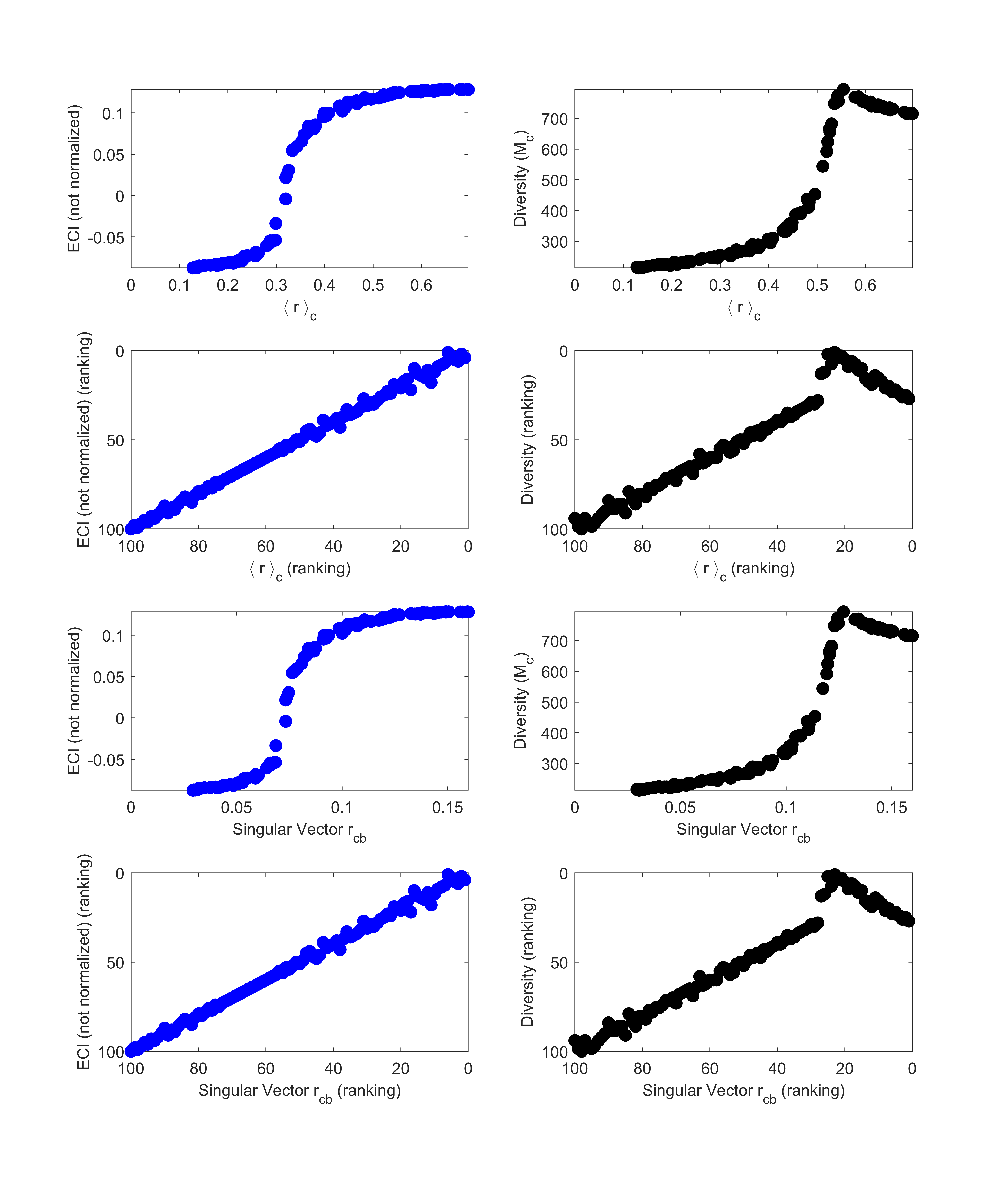}
    \caption{Relationship between the average capability level of an economy $\langle r \rangle_c$, the leading singular vector of the capability matrix $r_{cb}$, $ECI$, and diversity, including their rankings. The model assumes that the probability an economy has a capability or that an activity requires it is composed of a structured component ($3/4$ linearly spaced in $[0,1]$) and a random component ($1/4$). The top two rows show the relationship between $\langle r \rangle_c$, $ECI$, and diversity, while the bottom two rows show the same relationships using the leading singular vector of $r_{cb}$ instead of the average.}
    \label{eigendiffcap}
\end{figure}

We can see that despite introducing substantial variation in the probability for a capability to be present, the matrices retain a similar shape. In fact, we find that $ECI$ continues to perform well as an estimator of both the average level of capabilities $\langle r \rangle_c$ and the leading singular vector of the capability matrix, as shown in Figure~\ref{eigendiffcap}. This means that in the context of a model with multiple capabilities we can interpret $ECI$ as an estimate of both the average capability level of an economy and of the dominant direction of variation in capabilities across locations.\\

\begin{figure}[htbp]
    \centering
    \includegraphics[width=0.9\textwidth]{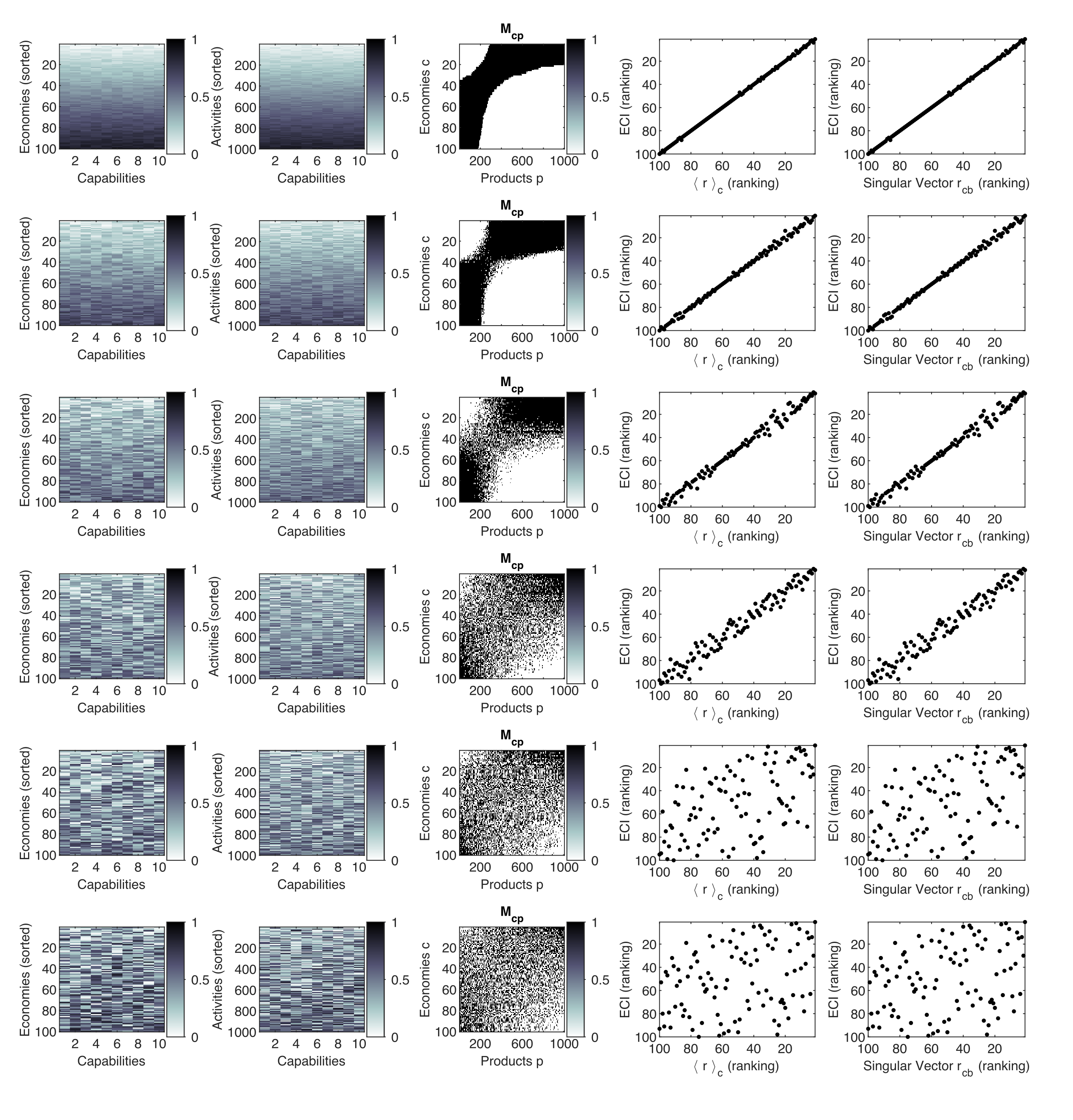}
    \caption{Numerical implementation of the multi-capability model for 100 economies, 1000 activities, and 10 capabilities. The capability probabilities of economies, or the probabilities that activities require them, follows the parametrization in equation (\ref{eq:parametrization}). Each row of this figure represents a different level of mixing between a baseline probability and a uniform random number. From top to bottom, the weight of the baseline $\alpha$ are 0.9, 0.75, 0.6, 0.45, 0.3, and 0.15.}
    \label{fig:multigrid}
\end{figure}

\begin{figure}[htbp]
    \centering
    \includegraphics[width=0.8\textwidth]{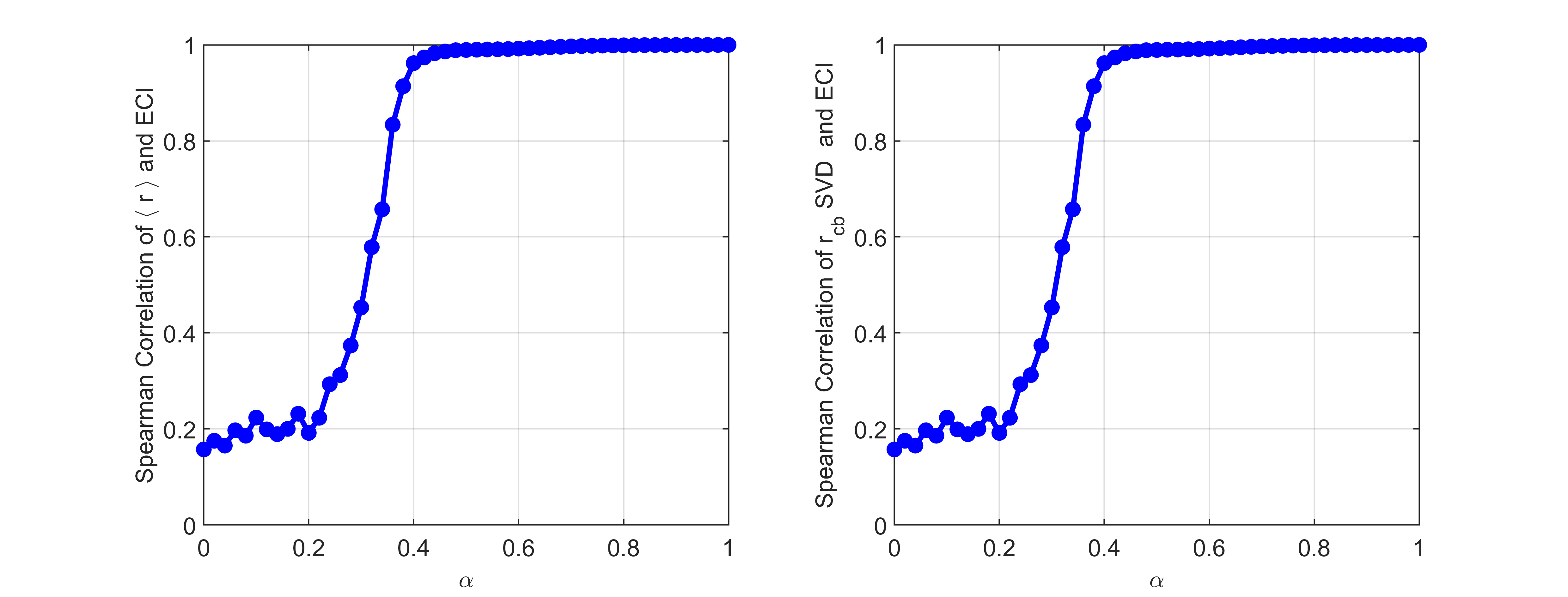}
    \caption{Average correlation between $ECI$ and 1) the average capability level of an economy $\langle r \rangle_c$ and 2) the leading singular vector or $r_{cb}$ as a function of the mixing probability $\alpha$. We can see that the correlation remains close to one for mixing probabilities above 0.4.}
    \label{fig:phase}
\end{figure}

But how far can we take this intuition? Does this method work for completely random distribution of capabilities? Or does it require an adequate level of correlation between the different capabilities?\\

We can explore this by using the parametrization introduced in equation~(\ref{eq:parametrization}) to vary the level of randomness in probability for the presence of capabilities. Figure~\ref{fig:multigrid} performs this exploration, by showing the capability presence matrices, $M_{cp}$, and the correlation between the ranks of $ECI$ and 1) the average capability level of an economy $\langle r\rangle_c$ and 2) the leading singular vector, for $\alpha=[0.9,0.75,0.6,0.45,0.3,0.15]$. This exercise reveals that $ECI$ is rather robust, and is able to capture both the average capability level and  the dominant pattern of variation for an economy even when the level is 60 percent random and 40 percent based on a baseline. This exercise also shows that the relationship between $ECI$ and both $\langle r\rangle_c$ and the first singular vector breaks somewhere between $\alpha=0.45$ and $\alpha=0.3$, suggesting a potential phase transition in this behavior. \\

Figure~\ref{fig:phase} explores this phase transition by presenting the average correlation between $ECI$ and $\langle r\rangle_c$ (left panel) and $ECI$ and the first singular vector (right panel) observed after sweeping through the parametrization parameter $\alpha$ 250 times using a linearly space grid of 50 points for the interval $\alpha=[0.01,1]$. We can see that there is a phase transition around 0.35, meaning that the ability of $ECI$ to recover the average capability level and the pattern of variation in $_rcb$ of an economy in this model ($\langle r\rangle_c$) is valid as long as there is a strong enough correlation among the capability probabilities of an economy.\\

Overall, despite the added complexity of the multi-capability model, and the added variation of using randomly drawn probabilities for $r_{cb}$ and $q_{pb}$, the behavior of the second eigenvector of $M_{cc'}$ or $ECI$, and the shapes of the matrices leading to its calculation, are largely consistent with the intuition we developed in the single capability model. That extends the robustness of this idea to models with a wide range of capabilities, including models with heterogeneous capabilities and substantial levels of noise on how those capabilities are assigned.\\

But are these observations particular to models based on capabilities and probabilities? Or can we use the second eigenvector method of economic complexity to recover factors in models based on other production functions?\\

In the next section we explore extensions of this method to other production functions to delineate the effective boundaries of this theory.\\

\section{Other Production Functions}
\label{sec:other_production_functions}

You may now be wondering whether the ability of the economic complexity index ($ECI$) to recover latent characteristics of economies is specific to stochastic capability models, or whether it extends to a broader class of production functions. A closely related question, emphasized in recent theoretical work, is whether this ability can be understood through the lens of log-supermodularity, which has been shown to generate sharp sorting and ranking results in assignment models of production~\cite{costinot2009origins,schetter2022measure,yildirim_sorting_2021}. In this section, we clarify the relationship between these perspectives and identify which properties of a production function matter for the construction and interpretation of $ECI$.\\

The main message is simple. Log-supermodularity is a useful property because it describes how production levels are sorted across economies and activities. But $ECI$ is not constructed directly from production levels. It is constructed from revealed comparative advantage ($RCA$), and then from the binary specialization matrix $M_{cp}$. These transformations can change the relevant structure. As a result, log-supermodularity is not necessary by itself, for $ECI$ to recover latent economy rankings. What matters is whether the production function generates an informative specialization matrix after the $RCA$ transformation.\\

To make this point in a unified way, consider the following CES family of factor-ratio production functions with a normalized baseline component:
\begin{equation}
    Y_{cp}=A\left[\delta\!\left(\frac{K_c}{K_p}\right)^{\!\rho}+(1-\delta)\right]^{1/\rho}.
    \label{eq:ces-v2}
\end{equation}
Here $K_c$ denotes the relevant factor level of economy $c$, while $K_p$ denotes the factor intensity requirement of activity $p$. A higher $K_c$ makes production easier, while a higher $K_p$ makes production more demanding. The parameter $\delta\in(0,1)$ is a distribution parameter, and $\rho$ governs the shape of the function.\\

It is important to interpret equation~\eqref{eq:ces-v2} carefully. The second term, $(1-\delta)$, is a normalized baseline component. Therefore, the function should not be read as a technology in which $K_c$ is the only necessary input. For $\rho>0$, this baseline implies that output may remain positive even when $K_c$ is very small. For $\rho<0$, the function behaves more like a bottleneck technology, with output approaching zero as $K_c$ approaches zero. We use this family not because every value of $\rho$ has the same economic interpretation, but because it allows us to compare separable, additive, and bottleneck-like cases within a common framework.\\

This family nests a Cobb--Douglas factor-ratio form as $\rho\to 0$:
\begin{equation}
    Y_{cp}=A\left(\frac{K_c}{K_p}\right)^{\delta}.
    \label{cobb-intensity-v2}
\end{equation}
It also converges to the Leontief factor-ratio production function as $\rho\to-\infty$
\begin{equation}
    Y_{cp}=A\min\left\{\frac{K_c}{K_p},1\right\}.
\end{equation}
The Leontief case has a natural bottleneck interpretation: an economy can fully produce an activity only if its factor level is high enough relative to the activity's requirement.\\

We first examine how this family relates to log-supermodularity. Since output depends on $K_c$ and $K_p$ only through the ratio $K_c/K_p$, the relevant question is whether the relative advantage of a high-$K_c$ economy is larger in high-$K_p$ activities. In other words, if $K_{c'}>K_c$ and $K_{p'}>K_p$, log-supermodularity requires
\begin{equation}
\frac{Y_{c'p'}}{Y_{c'p}}>\frac{Y_{cp'}}{Y_{cp}}.
\label{eq:log-super-discrete-v2}
\end{equation}
This means that when we move from a less demanding activity $p$ to a more demanding activity $p'$, the proportional change in output is more favorable for the high-capability economy than for the low-capability economy.\\

For the CES family in equation~\eqref{eq:ces-v2}, the corresponding continuous condition is obtained from the cross-partial derivative of $\ln Y_{cp}$ with respect to $\ln K_c$ and $\ln K_p$:
\begin{equation}
\frac{\partial^2\ln Y_{cp}}{\partial\ln K_c\,\partial\ln K_p}
= -\frac{\rho\,\delta(1-\delta)(K_c/K_p)^{\rho}}
{\bigl[\delta(K_c/K_p)^{\rho}+(1-\delta)\bigr]^{2}}.
\label{eq:ces-cross-v2}
\end{equation}
The sign of this expression is determined by $-\rho$. Thus, the production function is log-supermodular when $\rho<0$, log-modular in the Cobb--Douglas limit $\rho=0$, and log-submodular when $\rho>0$. Economically, the complementary case $\rho<0$ gives the usual positive sorting intuition: high-$K_c$ economies have a relative advantage in more demanding activities. The substitutable case $\rho>0$ reverses this sorting because the baseline component dampens the proportional disadvantage of low-$K_c$ economies in demanding activities.

The Cobb--Douglas case illustrates an important point. In the limit $\rho\to 0$, output is
\begin{equation}
    Y_{cp}=A\left(\frac{K_c}{K_p}\right)^{\delta}
    =
    A K_c^\delta K_p^{-\delta}.
\end{equation}
This is multiplicatively separable in an economy component and an activity component. Applying Balassa's (1965) revealed comparative advantage transformation,
\begin{equation}
    R_{cp}=\frac{Y_{cp}\sum_{c,p}Y_{cp}}{\sum_p Y_{cp}\sum_c Y_{cp}},
\end{equation}
all multiplicative terms cancel out, giving
\begin{equation}
    R_{cp}=1
    \quad \forall c,p.
\end{equation}
Thus, even though production levels are ordered by $K_c$ and $K_p$, there is no revealed specialization. Every economy has the same revealed comparative advantage in every activity. The $RCA$ matrix therefore contains no information from which $ECI$ could recover latent economy rankings.

This result is not specific to Cobb--Douglas. It applies to any multiplicatively separable production function of the form:\\
\begin{equation}
    Y_{cp}=A f(K_c) g(K_p).
    \label{sep-v2}
\end{equation}\\

For such functions, the economy component and the activity component cancel mechanically under the $RCA$ normalization. Hence, since it makes all economies look equally specialized in all activities, multiplicative separability is fatal for $ECI$.\\

This observation helps separate three ideas that are often conflated.\\

First, non-separability is necessary for informative specialization. If production is multiplicatively separable, the $RCA$ matrix is trivial.\\

Second, log-supermodularity describes sorting in production levels, not necessarily sorting in the $RCA$ matrix. In a log-submodular function high-complexity economies will specialize in low complexity activities because they will produce an enormous amount of output in them, even though they will still produce more output in high-complexity activities than low-complexity economies. In that case, the matrices will also cluster low- and high- complexity economies, which is the condition needed to estimate economic complexity. So economic complexity can also be recovered for log-submodular functions, even if the world does not behave that way.\\ 

What $ECI$ needs is a specialization matrix with enough structure for the eigenvector to recover the latent ordering of economies. The examples below show that this can happen with log-supermodular, log-submodular, and non-additive technologies, but through different mechanisms.\\

A useful sufficient condition is the additive form:
\begin{equation}
    Y_{cp} = B_c + f_c\,g_p,
    \label{shifted-v2}
\end{equation}
where $B_c\geq 0$ is an economy-specific baseline (with a strict inequality for some $c$'s), $f_c$ captures the economy's productive capacity, and $g_p$ captures the activity characteristic relevant for specialization. The baseline term $B_c$ is what breaks multiplicative separability. The single-capability model is a special case with $B_c=B=1$, $f_c=r_c-1$, and $g_p=q_p$.\\

The advantage of this form is that the $RCA$ condition can be written in a simple way.\\

Define:\\
\[
G=\sum_p g_p, \qquad B=\sum_c B_c, \qquad F=\sum_c f_c.
\]
The marginal totals are:\\
\[
Y_c = N_p B_c + f_c G, 
\qquad 
Y_p = B + Fg_p, 
\qquad 
Y = N_p B + FG.
\]
The condition $R_{cp}\geq 1$ is equivalent to
\[
Y_{cp}Y\geq Y_cY_p.
\]
Substituting equation~\eqref{shifted-v2} and simplifying gives
\[
(G - N_p g_p)(F B_c - B f_c)\geq 0.
\]
Equivalently,
\begin{align}
   \left(g_p-\langle g\rangle\right)
    \left(\langle B\rangle f_c-B_c\langle f\rangle \right)
    \geq 0,
    \label{eq:gc-v2}
\end{align}

This condition has a transparent interpretation. Economies are ranked by:\\
\[
\langle B\rangle f_c-B_c\langle f\rangle,
\]
while activities are ranked by:\\
\[
g_p-\langle g\rangle.
\]
An economy has revealed comparative advantage when the two terms have the same sign. Therefore, economies above the relevant threshold specialize in activities with above-average $g_p$, while economies below the threshold specialize in activities with below-average $g_p$. This generates a block structure in the specialization matrix:\\
\begin{equation}
\begin{aligned}
M_{cp} &= 1, \quad \text{if} \quad 
\frac{f_c}{\langle f\rangle}\geq\frac{B_c}{\langle B\rangle} 
\ \text{and}\ 
g_p \geq \langle g \rangle,\\
M_{cp} &= 1, \quad \text{if} \quad 
\frac{f_c}{\langle f\rangle}<\frac{B_c}{\langle B\rangle} 
\ \text{and}\ 
g_p < \langle g \rangle,\\
M_{cp} &= 0, \quad \text{otherwise}.
\end{aligned}
\end{equation}

When the baseline is common across economies, $B_c=B_0$ for all $c$, the condition simplifies further:\\
\begin{equation}
(f_c-\langle f\rangle)(g_p-\langle g\rangle)\geq 0.
\label{eq:gc-constant-v2}
\end{equation}
In this case, $ECI$ ranks economies by $f_c$. This is the cleanest case: the additive production function generates a block-structured specialization matrix, and the eigenvector recovers the latent productive capacity.\\

We now illustrate this result with two examples. \\

The first example is the linear CES case, obtained by setting $\rho=1$ in equation~\eqref{eq:ces-v2}:
\[
Y_{cp}=A\!\left[\delta\frac{K_c}{K_p}+(1-\delta)\right].
\]
This case is useful precisely because it is not log-supermodular. It is log-submodular. A high-$K_c$ economy has a larger relative output advantage in less demanding activities than in more demanding activities. This happens because the baseline term $A(1-\delta)$ protects low-$K_c$ economies from falling too much when $K_p$ rises. Thus, in terms of production levels, the sorting pattern is the reverse of the standard log-supermodular case.\\

Nevertheless, the $RCA$ matrix remains informative. The reason is that the linear CES can be rewritten as
\[
Y_{cp}
=
\underbrace{A(1-\delta)}_{B}
+
\underbrace{A\delta K_c}_{f_c}
\cdot
\underbrace{\frac{1}{K_p}}_{g_p}.
\]
This is exactly of the additive form in equation~\eqref{shifted-v2}, with a common baseline $B=A(1-\delta)$, economy term $f_c=A\delta K_c$, and activity term $g_p=1/K_p$. Since the baseline is common across economies, condition~\eqref{eq:gc-constant-v2} applies. Substituting the terms gives
\[
\left(K_c-\langle K\rangle_c\right)
\left(\frac{1}{K_p}-\left\langle\frac{1}{K}\right\rangle_p\right)
\geq 0.
\]
Thus, high-$K_c$ economies have revealed comparative advantage in activities with high $1/K_p$, that is, in less factor-intensive activities. This is a reversed form of specialization relative to the usual capability-complexity interpretation, but it is still structured. The eigenvector can still recover the latent ordering of economies by $K_c$ as high complexity economies will be sorter in one group and low complexity in another. Thus, this example shows why log-supermodularity is not necessary for $ECI$ to recover an economy ranking. The sign of the cross-partial in~\eqref{eq:ces-cross-v2} determines the direction of sorting in production levels, while the additive form determines whether the $RCA$ matrix has a clean block structure.\\

The second example shows how the condition works when the baseline varies across economies. Consider an extension of the single-capability model in which the capability level $r_c$ also scales output directly:
\begin{equation}
    Y_{cp} = r_c^\alpha\bigl(1-q_p(1-r_c)\bigr), 
    \qquad \alpha\geq 0.
    \label{eq:rcalpha-v2}
\end{equation}
For $\alpha=0$, this reduces to the original single-capability model. Expanding the expression gives\\
\[
Y_{cp}
=
\underbrace{r_c^\alpha}_{B_c}
+
\underbrace{r_c^\alpha(1-r_c)}_{f_c}
\cdot
\underbrace{(-q_p)}_{g_p}.
\]\\
This is again of the additive form in equation~\eqref{shifted-v2}, but now the baseline $B_c=r_c^\alpha$ varies across economies.\\

Substituting these terms into condition~\eqref{eq:gc-v2} gives\\
\begin{equation}
(q_p-\langle q\rangle)
\left(
r_c-\frac{\langle r^{\alpha+1}\rangle}{\langle r^\alpha\rangle}
\right)
\geq 0.
\label{eq:rcalpha-v2-cond}
\end{equation}\\
The block structure is preserved. The only difference from the original single-capability case is that the threshold is no longer $\langle r\rangle$. Here, the threshold is a weighted average of $r_c$, with greater weight placed on high-$r_c$ economies when $\alpha>0$. Hence, the threshold rises as the direct productivity role of $r_c$ becomes stronger. Intuitively, comparative advantage in complex activities becomes concentrated among a smaller set of high-capability economies.\\

The additive form in equation~\eqref{shifted-v2} is therefore sufficient for $ECI$ to recover a latent ranking, but it is not necessary. To see this, consider the Leontief limit of the CES family:\\
\begin{equation}
Y_{cp}=A\min\left\{\frac{K_c}{K_p},1\right\}.
\end{equation}
This production function is log-supermodular. High-$K_c$ economies have a relative advantage in high-$K_p$ activities because they can fully produce activities that low-$K_c$ economies cannot. Yet the Leontief function is not of the additive form in equation~\eqref{shifted-v2}.\\

The reason is intuitive. In the additive form, differences across activities are proportional to a common economy-specific factor $f_c$. In the Leontief case, however, output is flat for all activities with $K_p\leq K_c$ and then declines once $K_p>K_c$. The location of this kink depends on the economy's own factor level $K_c$. Because each economy has a different saturation point, the production surface cannot be represented as a common baseline plus a product of one economy term and one activity term.\\

This also changes the shape of the specialization matrix. The additive case gives a block structure leading to economies above a threshold to specialize in one group of activities, while economies below it specialize in another. The Leontief case instead gives a staircase structure. When economies are ordered by $K_c$ and activities by $K_p$, each economy has an advantage in activities up to a capability-dependent boundary. Higher-$K_c$ economies can sustain comparative advantage in progressively more demanding activities because they will produce the maximum available production ($Y_{cp} = 1$). This means that the binary specialization matrix will resemble\\
\[
M_{cp}=\mathbf{1}[K_c\geq K_p].
\]\\
The diversity of economy $c$ is then the number of activities for which $K_p\leq K_c$,
which is monotone in $K_c$. Therefore, $ECI$ can still recover the latent ordering, but now it does so through a diversity-based staircase rather than through the two-block structure generated by the additive class.\\

This example is important because it shows that the additive form is not necessary. A production function might not belong to the additive class and still generate an $M_{cp}$ matrix from which $ECI$ recovers the latent ranking. The mechanism, however, is different. In the additive case, $ECI$ works because specialization divides economies and activities into blocks. In the Leontief case, $ECI$ works because diversity is monotone in capability.\\

Together, these examples clarify the relationship between production functions, log-supermodularity, and economic complexity. The Cobb--Douglas case is log-modular and multiplicatively separable, so it fails: all $RCA$ values are equal to one, and there is no specialization structure for $ECI$ to use. The linear CES case is log-submodular, yet it belongs to the additive class and generates a structured $RCA$ matrix from which $ECI$ can recover the ordering of economies. The Leontief case is log-supermodular and does not belong to the additive class, but it can still allow $ECI$ to recover the latent ordering through a staircase specialization pattern.\\

The broader lesson is that log-supermodularity is not necessary for $ECI$ to recover latent capability rankings. It is a property of production levels, while $ECI$ is constructed from normalized and discretized specialization patterns. The essential question is not only whether production levels exhibit positive sorting, but whether the $RCA$ transformation produces a matrix with a recoverable ordering.\\

Finally, the criterion of whether $ECI$ recovers latent capability rankings from a given production function carries an empirical implication about the nature of real production. If real production were well described by a multiplicatively separable function such as Cobb–Douglas, the RCA matrix would be uniformly uninformative and $ECI$ would have no structure to exploit. Yet, empirically, $ECI$ does recover meaningful rankings. This suggests that real production is not multiplicatively separable. Conversely, if real production was a pure Leontief threshold, $ECI$ would reduce to a diversity ordering. The empirical fact that $ECI$ outperforms raw diversity as a predictor of economic growth suggests real specialization matrices are not pure staircases. That is, the data appear to contain richer patterns of specialization, closer to block or multi-factor structures than to a simple one-dimensional threshold model. This is consistent with real production requiring multiple capabilities and heterogeneous activity requirements rather than a single scalar factor alone.\\

\section{Uniqueness of $ECI$}

The previous sections provide a principled way to interpret empirical metrics of economic complexity, such as the $ECI$. In a nutshell, we can say that \textbf{a metric of economic complexity $K_c$ is a ``good'' metric if it is a strictly monotonic and invertible function of $r_c$,} so that it preserves the ordering and allows us to recover the underlying capability level. Yet this definition is not unique to $ECI$: in principle, other transformations of observed outcomes could also be strictly monotone in $r_c$. For instance, could we recover $r_c$ directly from $Y_{cp}$ or from $R_{cp}$? If we can, then why bother with $ECI$? A key point is that the structural production mapping can satisfy log-supermodularity -- so that higher-capability economies sort into more capability-intensive activities in a way that supports separability -- while the empirical problem of estimating $r_c$ from observed matrices is not a direct test of log-supermodularity; rather, it is an eigenvector-recovery problem built from noisy and heterogeneously scaled observations, and these features can scramble rankings, inflate or deflate correlations, and interact with scale factors in ways that make naive eigenvectors of $YY^\top$ or $RR^\top$ poor proxies for the latent capability ordering even when the underlying mapping remains log-supermodular.

\begin{figure}[htbp]
\centering
\includegraphics[width=0.95\textwidth]{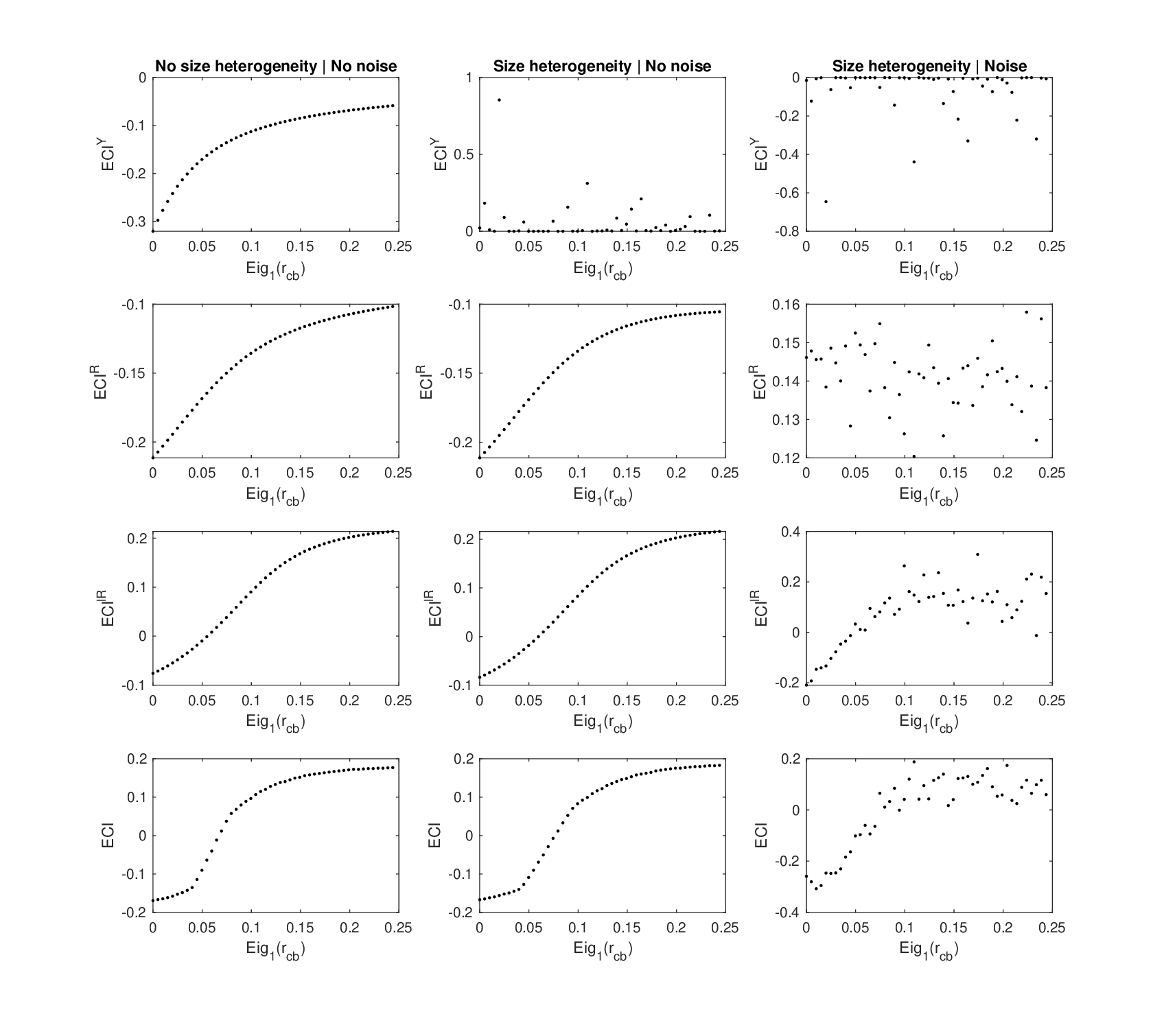}
\caption{Comparison between four methods to estimate economic complexity ($ECI^Y$, $ECI^R$, $ECI^{lR}$, and $ECI$) for production functions including no size heterogeneity nor noise (column 1), size heterogeneity but no noise (column 2), and size heterogeneity and noise (column 3). Size heterogeneity was implemented by setting $A_c=\textrm{exp}(10\times\textrm{rand}(0,1))$ and $A_p=\textrm{exp}(10\times\textrm{rand}(0,1))$ and noise in column 3 is set at $\alpha=0.7$ (that is 30 percent of output is noise.)}
\label{fig:whyeci}
\end{figure}

In this section we explore alternative metrics to clarify when $ECI$ is warranted and why. The lesson is not that noise or scale necessarily destroys log-supermodularity of the structural mapping, but that empirical recovery of $r_c$ from observed output requires robustness to two sources of variation that are largely orthogonal to the log-supermodularity structure: heterogeneity in the size of units of observation and noise in measured outcomes. In practice, these features can reshuffle empirical rankings in finite samples, distort the cross-country similarity structure by inflating or deflating correlations, and interact with the scale factors that load into $Y_{cp}$ multiplicatively, so that eigenvectors formed from $YY^\top$ or even from $RR^\top$ need not align with the latent capability ordering even when the underlying mapping is log-supermodularity.\\

Let us start with our basic model:\\
\begin{equation}
Y_{cp}=A\prod_{b}(1-q_{pb}(1-r_{cb})).
\end{equation}\\

In this model, the output of an economy in an activity $Y_{cp}$ depends only on the availability of capabilities. This provides a clean benchmark in which the mapping from capability levels to output is log-supermodular in the relevant arguments, so that comparative advantage sorts higher-$r_c$ economies toward more capability-demanding activities. Yet this is a crude oversimplification since in economic geography, units of observation are heterogeneous. For example, the GDP of the United States (at nearly 30T USD in 2005) is more than 1,000 times that of Mozambique. Similarly, crude Petroleum exports (usually over 1T a year) are about 20,000 times larger than those of ``Wool Hair Waste.''\footnote{These heterogeneities are also largely exogenous, since international borders are the result of historical processes.} A more reasonable model therefore incorporates the sizes of economies and activities through scale factors:

\begin{equation}
Y_{cp}=A_cA_p\prod_{b}(1-q_{pb}(1-r_{cb})),
\end{equation}\\

where $A_c$ is a scale factor for each economy and $A_p$ is a scale factor for each activity (e.g. the size of that activity's global market).

Size differences mean that we cannot simply associate output to capabilities, since economies that are larger (for historical reasons) can produce more total output than smaller economies even when they have lower capability levels. For example, India exports more pharmaceuticals than Denmark, but with a population and GDP that are orders of magnitude larger than Denmark's. This observation motivates the use of specialization metrics, such as RCA, in empirical research: they are designed to factor out scale and to better isolate the comparative-advantage sorting implied by the log-supermodularity structure of the production component, yielding an estimate of output that controls for heterogeneity in the sizes of units of observation.\\

Finally, we can make the estimation problem more realistic by adding noise to the production function. That is, we let:

\begin{equation}
Y_{cp}=A_cA_p\big(\alpha\prod_{b}(1-q_{pb}(1-r_{cb}))+(1-\alpha)\text{rand}(0,1)\big),
\end{equation}\\

where $\alpha$ is a tunable noise parameter $[0,1]$ and $\text{rand}(0,1)$ is a random number between $0$ and $1$. This form of noise captures the idea that administrative data are imperfect and that outcomes are affected by idiosyncratic short-term shocks. Importantly, even if the underlying production function remains super-modular, the empirical task is to recover $r_c$ from matrices built from noisy $Y_{cp}$ across many $c,p$, and noise can reshuffle observed rankings in finite samples, distort similarity patterns, and interact with the multiplicative scale factors $A_cA_p$, making eigenvectors of $YY^\top$ or $RR^\top$ unreliable proxies for latent capability levels unless the metric is designed to be robust to these distortions.\\

Building on this setup, we explore the ability of a few different measures of economic complexity to recover the first eigenvector of the capability level matrix: $r_{cb}$. In addition to $ECI$, we consider the following three alternatives:

\begin{equation}
\begin{aligned}
    ECI^{Y}&=\textrm{first eig. of } M_{cc'}=\sum_pY_{cp}Y_{c'p},\\
    ECI^{R}&=\textrm{first eig. of } M_{cc'}=\sum_pR_{cp}R_{c'p},\\
    ECI^{lR}&=\textrm{first eig. of } M_{cc'}=\sum_p \textrm{log}(R_{cp}+0.1)\textrm{log}(R_{c'p}+0.1).
\end{aligned}
\end{equation}

The first one ($ECI^Y$) asks whether we can obtain a meaningful measure of economic complexity directly from the output matrix (made square by multiplying it by its transpose). The second one ($ECI^R$) asks whether we can obtain a good measure of complexity from the specialization matrix $R$, which is designed to remove scale heterogeneity. The last one ($ECI^{lR}$) asks whether we can obtain a good metric of economic complexity from the specialization matrix $R$ once we transform it further, reducing its variance with a logarithmic transformation and capping small values by adding $1/10$.\\

Figure~\ref{fig:whyeci} shows the relationship between these four measures of complexity and the first eigenvector of the capability matrix for the three production functions in this section.\\

The first column shows the case with no size heterogeneity and no noise. In this case, all of the proposed metrics of economic complexity work, providing an injective, invertible, and strictly monotonic transformation of the first eigenvector of $r_{cb}$.\\

The second column shows the impact of size heterogeneity. In this case, the first eigenvector of the capability matrix cannot be recovered from the output matrix, but it can be recovered from the specialization matrix $R$ and $ECI$.\\

The last column shows an example where the output is a mixture involving 30 percent noise. In this case, the eigenvectors of neither the output matrix nor the specialization matrix provides a good approximation of the capability levels. Yet, both $ECI$ and the eigenvector of $\textrm{log}(R+0.1)$ still recover the capability level matrix, albeit with some noise.\\

The results are summarized also in Table~\ref{table:whyeci}. In sum, the use of the economic complexity index as a method estimate capability levels, can be justified in a world with heterogeneous units of observation and noise. In this world, also a reasonable measure of economic complexity can be obtained by taking the eigenvector of the logarithm of the specialization matrix. The take home message is that, in these more realistic settings, the recovery of the capability levels requires transformations to the output matrix that go beyond the estimation of specialization. So we can think of metrics of economic complexity in this context, as those that provide estimates of capability levels in a way that is robust to size heterogeneity and the presence of noise.

\begin{center}
\renewcommand{\arraystretch}{1.5}
\begin{tabular}{|l|c|c|c|c|}
\hline
\textbf{Production function} &  $ECI^Y$  & $ECI^R$  & $ECI^{lR}$ & $ECI$\\ \hline
Homogeneous Size $\&$ No Noise & \Checkmark  & \Checkmark  & \Checkmark & \Checkmark \\ \hline
Heterogeneous Size $\&$ No Noise & \XSolidBrush & \Checkmark & \Checkmark & \Checkmark\\ \hline
Heterogeneous Size $\&$ Noise  & \XSolidBrush & \XSolidBrush & \Checkmark & \Checkmark \\ 
\hline
\end{tabular}
\label{table:whyeci}
\end{center}

\section{Prices, Wages, and Consumption}
\label{sec:prices-wages-consumption}
We conclude our theoretical exploration by considering an extension of the single-capability model to a short-run equilibrium framework, with variable prices, wages, and consumption. This extension helps connect the economic complexity framework with traditional modeling approaches, helping provide a linchpin connecting the multiple areas of research that are associated with work in economic complexity, such as evolutionary economics, economic development, and economic geography.\\

We let the output of an economy in an activity depend explicitly on the price of each activity $\pi_p$ by generalizing our output function to:\\

\begin{equation}
    Y_{cp}=\pi_p(1-q_p(1-r_c))=\pi_py_{cp}.
\end{equation}

We use this function to explore a few things. First, we derive a simple relationship between capability levels and wages. Then, we derive a new condition from the specialization matrix $R_{cp}$, which is the key condition connecting the empirical economic complexity estimate $ECI$ with the model's capability level. Finally, we estimate product prices by exploring an extension of the model where economies maximize their utility of consumption constrained by their income and the global supply of goods.\\

First, we focus on wages.\\

In a perfectly competitive market where labor is the only factor, and all income goes into wages, then the total income of an economy $Y_c$ must equal the wages $w_c$ it pays times the amount of labor $L_c$ it employs. That is:\\

\begin{equation}
    Y_{c}=w_cL_c,
\end{equation}

which implies:

\begin{equation}
    w_{c}=\frac{\sum_p\pi_p(1-q_p(1-r_c))}{L_c}.
\end{equation}

Dividing the numerator and denominator by $1/N_p$ (one over the total number of activities), we can transform the sums into averages to obtain an expected wage $w_c^*$:

\begin{equation}
    w^*_{c}=\frac{N_p\left(\langle\pi\rangle+\langle q\pi\rangle(r_c-1)\right)}{L_c}.
    \label{wage}
\end{equation}

The above equation implies that wages are proportional to the capability level $r_c$, which we interpret as a measure of human capital, knowledge, or skill in that capability. In fact, wages grow in proportion to the product of prices times the probability an activity requires a capability and are inversely proportional to population:

\begin{equation}
    \frac{dw^*_{c}}{dr_c}=\frac{N_p\langle q\pi\rangle}{L_c}.
\end{equation}

This finding is consistent with the notion that economic complexity, which we now understand as an estimate of $r_c$, implies an equilibrium level of wages for an economy, and thus, explains future economic growth. In this model, economies must have a wage given by eqn.~\eqref{wage} in equilibrium. When out of equilibrium, economies should adjust (to first order) according to:

\begin{equation}
    \frac{dw_{c}}{dt}\propto -\eta(w_c-w_c^*),
\end{equation}

where $\eta$ is some proportionality constant (e.g. a speed or rate of adjustment). Economies with wages larger than equilibrium experience a downward pressure, whereas those with wages lower than equilibrium experience an upward pressure on their incomes.\\

Next, we calculate $R_{cp}$ to determine the condition separating the two specialization clusters that are key to determining economic complexity. Going back to the definition of $R_{cp}$ implies the condition :\\

\begin{equation}
    R_{cp}=\frac{\pi_p(1-q_p(1-r_c))\sum_{cp}\pi_p(1-q_p(1-r_c))}{\sum_{c}\pi_p(1-q_p(1-r_c))\sum_p\pi_p(1-q_p(1-r_c))}\geq 1,
\end{equation}\\

which after some algebra results in the inequality:\\

\begin{equation}
    (r_c-\langle r\rangle)(q_p\langle \pi \rangle-\langle q\pi\rangle)\geq 0.
\end{equation}\\

This brings us again to a specialization condition based on two clusters where economies with an above average probability of having the capability ($r_c>\langle r \rangle$) are specialized in products with a higher probability of requiring the capability, and where those with a below average probability of having the capability ($r_c<\langle r \rangle$) specialize in less demanding products. Yet, the threshold for activities is now:\\
\begin{equation}
    q_p \geq \frac{\langle q \pi\rangle}{\langle \pi \rangle},
\end{equation}\\
which, using the standard covariance identity\\
\begin{equation}
\langle q\pi\rangle =\langle q \rangle\langle\pi\rangle+\text{cov}(q,\pi),
\end{equation}
yields:\\
\begin{equation}
    q_p \geq \langle q \rangle + \frac{\text{cov}(q,\pi)}{\langle \pi \rangle}.
\end{equation}\\
This means that we recover the naked single-capability model when prices are uncorrelated with the probability that an activity requires a capability (when $\text{cov}(q,\pi)=0$). This equation also tells us that the specialization of high complexity economies in demanding activities is more pronounced when there is a positive correlation between the price of an activity and the probability it requires the capability in the model (which is a reasonable assumption). That is, in a world where prices are higher for more demanding activities, high complexity economies will specialize in a more narrow set of complex activities. Yet, for the purposes of this paper, what is important is that the specialization matrix is still divided into two clusters, just like in the single-capability model with no prices, and that these clusters separate among economies with high and low capability levels. \\

Finally, we explore an extension of this model including a demand side, by assuming a logarithmic utility function. That is, we let the utility of economy $c$ be given by:\\
\begin{equation}
U_c=\sum_p \theta_{cp}\log(C_{cp}).
\end{equation}\\
This assumption implies that agents experience a concave utility of consumption, meaning that they derive more utility from the first unit of consumption of a good or activity $p$ than from the last. Here $U_c$ should be interpreted as the utility of a representative consumer in economy $c$, with $C_{cp}$ denoting aggregate consumption of product $p$. Logarithmic preferences are convenient in this regard: they generate constant expenditure shares that are independent of income, which is sufficient for exact Gorman aggregation~\cite{gorman1953}. Aggregate demand therefore admits a representative-consumer representation regardless of how income is distributed across individuals, so the economy-level formulation above is well-defined even under heterogeneous consumers sharing common preferences $\theta_{cp}$.\\

We also assume that consumption is limited by the budget constraint:\\
\begin{equation}
\sum_p \pi_{p}C_{cp}\leq Y_c,
\end{equation}\\
which means that economies consumption is limited by the revenue generated by their total output. We also assume that the global production of goods is limited by the availability of capabilities, thus:\\ 
\begin{equation}
\sum_p C_{cp}=y_c.
\end{equation}\\
This means that in this model production capacity is fixed, and what adjusts is the price of an activity based on how demanding it is and how preferences for that activity are distributed across economies.\\ 

We start by maximizing utility following the Lagrangian:\\
\begin{equation}
\mathcal{L}=\sum_p \theta_{cp}\text{log}(C_{cp})-\lambda(\sum_p \pi_{p}C_{cp}- Y_c).
\end{equation}\\
Differentiating against consumption $C_{cp}$ and equating to zero we obtain the condition:\\
\begin{equation}
C_{cp}=\frac{\theta_{cp}}{\lambda\pi_{p}}.
\end{equation}\\
And using the budget constraint equation (which we use here as an equality) we can solve for $\lambda$:\\
\begin{equation}
\sum_p \pi_{p}\frac{\theta_{cp}}{\lambda\pi_{p}}= Y_c \quad \xrightarrow{} \lambda=\frac{\sum_p \theta_{cp}}{Y_c},
\end{equation}\\
meaning that consumption is given by:\\
\begin{equation}
C_{cp}=\frac{\theta_{cp}Y_c}{\pi_{p}\sum_{p'} \theta_{cp'}}.
\end{equation}
Finally, since:\\
\begin{equation}
Y_c=N_p (\langle \pi\rangle-(1-r_c)\langle q\pi\rangle),   
\end{equation}\\
consumption is given by:\\
\begin{equation}
C_{cp}=\frac{\theta_{cp}N_p (\langle \pi\rangle-(1-r_c)\langle q\pi\rangle)}{\pi_{p}\sum_p \theta_{cp}}.   
\end{equation}\\
Moving the $N_p$ to the denominator allows us to transform the remaining sum into an average:\\
\begin{equation}
C_{cp}=\frac{\theta_{cp} (\langle \pi\rangle-(1-r_c)\langle q\pi\rangle)}{\pi_{p} \langle \theta_{c}\rangle},  
\end{equation}\\
which means that consumption is downward slopping with the price of a good ($\pi_p$ appears only in the denominator \footnote{$\pi_p$ also appears implicitly in the average $\langle \pi\rangle$ and $\langle q\pi\rangle$ but its contribution is much smaller (divided by $1/N_p$). Also, the average can be thought of as a common price level, since it is the same for all products $p$}) and grows with an economy's preference for a specific activity ($\theta_{cp}$) and its capability level ($r_c$).\\

Now, we estimate prices by using the market clearing condition:\\
\begin{equation}
\sum_c C_{cp}=y_p,
\end{equation}\\
and since:\\
\begin{equation}
y_p=N_c(1-q_p(1-\langle r \rangle))
\end{equation}\\
it follows that:\\
\begin{equation}
\sum_c \frac{\theta_{cp} (\langle \pi\rangle-(1-r_c)\langle q\pi\rangle)}{\pi_{p} \langle \theta_{c}\rangle} =N_c(1-q_p(1-\langle r \rangle)).
\end{equation}\\
The above equation after some algebra can be brought to the form\footnote{here we used the notion that averages are constants to arrive to an expression where $\pi_p$ is expressed as a function of its ensemble averages.}:\\
\begin{equation}
\pi_p=\frac{\sum_c\frac{\theta_{cp}}{\langle \theta_c \rangle}(\langle\pi\rangle-\langle q\pi\rangle(1-r_c))}{N_c(1-q_p(1-\langle r \rangle))},
\end{equation}\\

which means the price of activity $p$ grows with the probability it requires the capability ($q_p$), since the denominator is the smallest it can be when $q_p=1$ and it is the maximized for $q_p=0$. Prices also grow when high capability economies (high $r_c$ and hence high-income $Y_c$ and high-wage economies) have a stronger preference ($\theta_{cp}$) for an activity.

\section{Relatedness and The Product Space}
\label{sec:relatedness}

Another observable used frequently in the economic complexity literature is the network of related activities \cite{hidalgo_product_2007,neffke_how_2011,jara-figueroa_role_2018,hidalgo_principle_2018,guevara_research_2016,alabdulkareem_unpacking_2018,knuepling2022does,klement2019innovation,stephany2024price,boschma_emergence_2013,cicerone_promoting_2020,chen_inter-industry_2017,ferrarini_product_2015,juhasz_explaining_2020,hidalgo_amenity_2020,borggren_knowledge_2016,muneepeerakul_urban_2013,chinazzi_mapping_2019,hausmann_atlas_2014}. When these activities are products, this network goes by the name of "product space." From an application perspective, the product space is used to estimate the potential of economy in an activity (e.g. the probability that a city specializes in an industry \cite{neffke_how_2011,neffke_skill_2013,jara-figueroa_role_2018,chen_inter-industry_2017}, a country starts exporting a product \cite{hidalgo_product_2007,hidalgo_principle_2018,hausmann_atlas_2014}, or a university starts producing papers in a given field \cite{guevara_research_2016,chinazzi_mapping_2019}. These estimates of potential are known as measures of relatedness, and are akin to traditional recommender system methods in computer science~\cite{resnick_recommender_1997}. Yet, in the economic complexity literature, they are used to explain economic development trajectories (e.g. countries entering new products) instead of individual consumption patterns (e.g. customers choosing to purchase a products at an online retailer) and to explore strategies to optimize industrial promotion efforts\cite{alshamsi_optimal_2018,waniek_computational_2020,stojkoski2025optimizing}\footnote{In recent years there have also been multiple efforts to look at relatedness in the context of sustainability, starting from the idea of a green product space, \cite{hamwey_mapping_2013,coniglio_production_2016,montresor_green_2019,mealy_economic_2017,huberty_green_2011,fraccascia_green_2018,perruchas_specialisation_2020,santoalha_diversifying_2020}}\\

Product space type networks are important in empirical work since they help capture information about an economy's productive structure that is specific to an economy and activity. Thus, they can be used to either model path dependencies, or to control for them in work looking at the impact of other factors in economic diversification\cite{boschma_institutions_2015,zhu_how_2017,huang_regional_2020,cortinovis_quality_2017}.\\

Here we begin by focusing on a particular characteristic of the product space that was emphasized when it was introduced nearly twenty years ago: the fact that the core of the product space, its most densely connected part, is composed of high-complexity activities\cite{hidalgo_product_2007}.\\

This is a characteristic that is particular to networks derived from trade data, since networks derived from other sources can have different forms. For example, networks connecting research fields based on citation patterns or co-authorships tend to follow a ``ring'' structure~\cite{guevara_research_2016,borner_design_2012}. Networks connecting skills based on the occupations that require them tend to follow a ``dumbbell'' structure (two big clusters connected by a bridge)~\cite{alabdulkareem_unpacking_2018}.\\

\begin{figure}[htbp]
    \centering
    \includegraphics[width=0.99\textwidth]{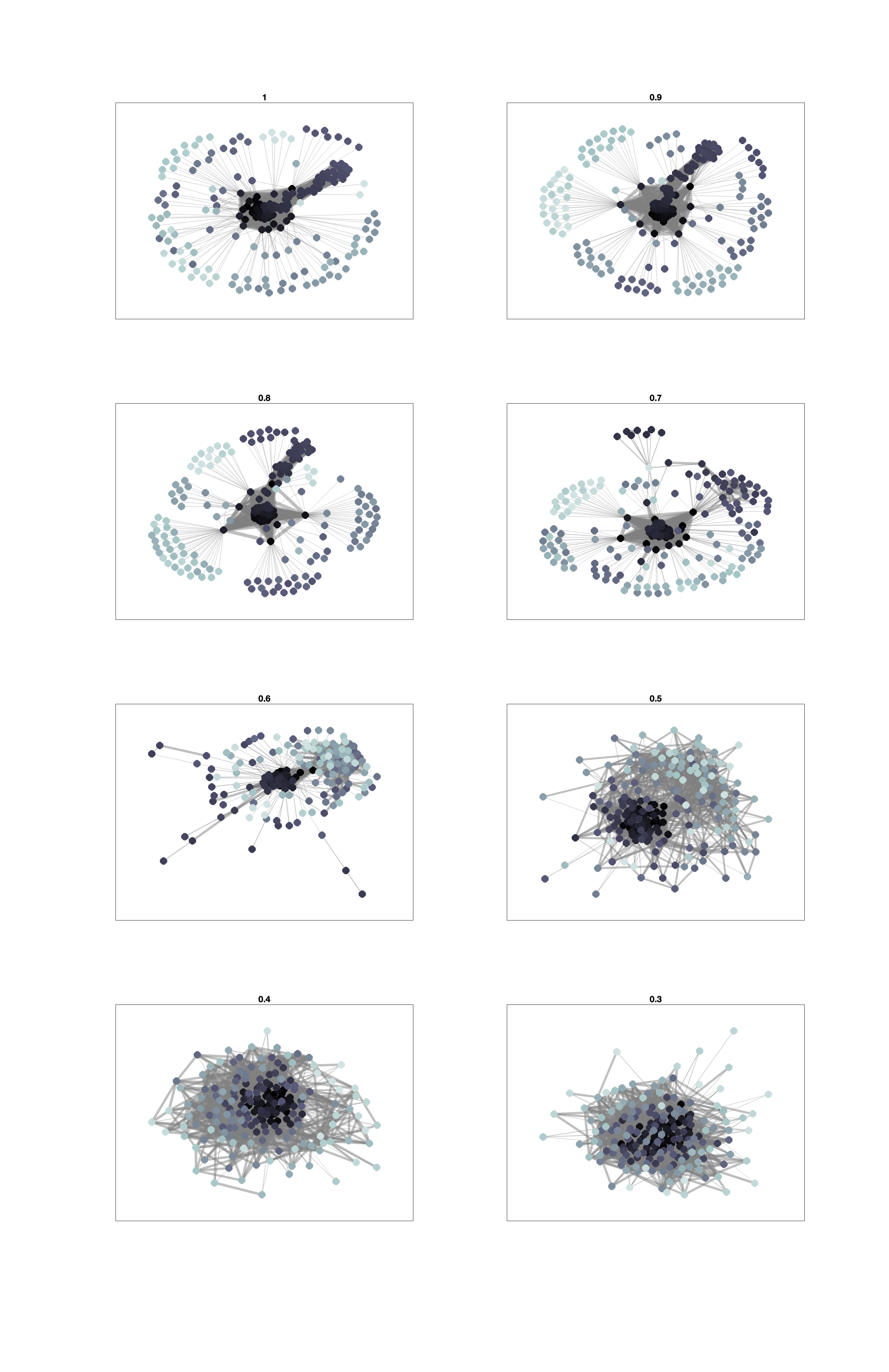}
    \caption{Networks of related activities generated for models with 200 activities, 100 economies, and 10 capabilities. Node shading corresponds to activity complexity (darker nodes denote greater complexity). Each panel is labeled with the value of the mixing parameter $\pi$, which specifies the degree to which random and non-random capabilities are combined (Equation~\eqref{eq:parametrization}).}
    \label{fig:net200p}
\end{figure}

We begin our exploration of the structure of the product space by estimating a measure of proximity, which is an estimate of the similarity between products. Unlike in the case of $ECI$, where we have a more strict definition based on a second eigenvector, measures of proximity, in both the economic complexity and recommender systems literature, tend to be more free. In \cite{hidalgo_product_2007} for instance, proximity was introduced using the minimum of the conditional probability that two products are exported in tandem. In our notation, this translates to:\\

\begin{equation}
\phi_{pp'}=\frac{\sum_c M_{cp}M_{cp'}}{\text{max}(M_p,M_{p'})}.
\end{equation}

In \cite{neffke_how_2011} proximity is simply the number of activities that are common to two economies.

\begin{equation}
\phi_{pp'}=\sum_c M_{cp}M_{cp'}.
\label{eq:simpleproximity}
\end{equation}

In general, it is not uncommon to find proximity matrices and recommender systems based on variations of $\sum_c M_{cp}M_{cp'}$ (usually with a normalization), so we will focus our exploration in this basic form.\\

The product space implied by the single-capability model can be derived easily for the case in which the number of economies and activities is even. In that case, the proximity matrices are:\\

\begin{align}
\phi_{pp'}=\sum_c M_{cp}M_{cp'}=M_p, \quad \text{if} \quad  q_p > \langle q \rangle \quad \& \quad q_{p'} > \langle q \rangle, \\
\phi_{pp'}=\frac{\sum_c M_{cp}M_{cp'}}{\text{max}(M_p,M_{p'})}=1, \quad \text{if}\quad  q_p > \langle q \rangle \quad \& \quad q_{p'} > \langle q \rangle.
\end{align}\\

This implies a network composed of two clusters, one connecting the activities that are produced in high-complexity economies, and one connecting the activities produced in low complexity economies.\\

A more interesting exercise is to consider the networks implied by the heterogeneous multi-capability model. Here we present three examples in which we estimate networks for different model parameters that we visualize by estimating their minimum spanning tree and adding on top of that all of the links that are one standard deviation above the mean. This is a similar visualization exercise than the one used in the paper that introduced the product space network and will ensure the same visualization method is used for all examples.\\

Figure~\ref{fig:net200p} presents this exercise for a model involving 200 activities, 100 economies, and 10 capabilities. Darker nodes indicate higher complexity. The number on top of each network visualization shows the mixing parameter $\alpha$ used to combine random and non-random capabilities.\\

We can see clearly in this example that all of the networks that are above the phase transition threshold are centered around a core of high-complexity activities, with lower complexity activities being peripheral. This reproduces the empirical observation that the core of the product space is composed of more sophisticated activities.\\ 

\begin{figure}[htbp]
    \centering
    \includegraphics[width=0.8\textwidth]{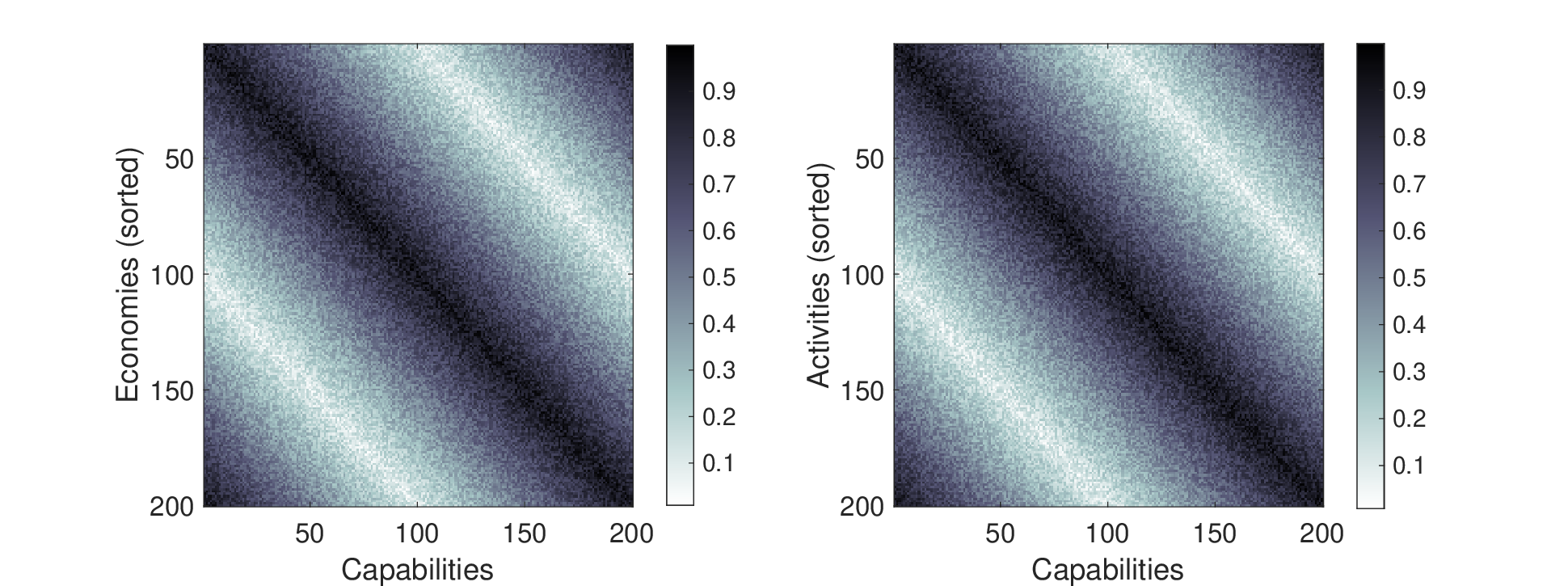}
    \caption{Parametrization of capability levels and requirements using a symmetric Toeplitz circulant matrix combined with a random matrix in 80 percent and 20 percent proportions.}
    \label{fig:toeplitzcap}
\end{figure}

\begin{figure}[htbp]
    \centering
    \includegraphics[width=0.7\textwidth]{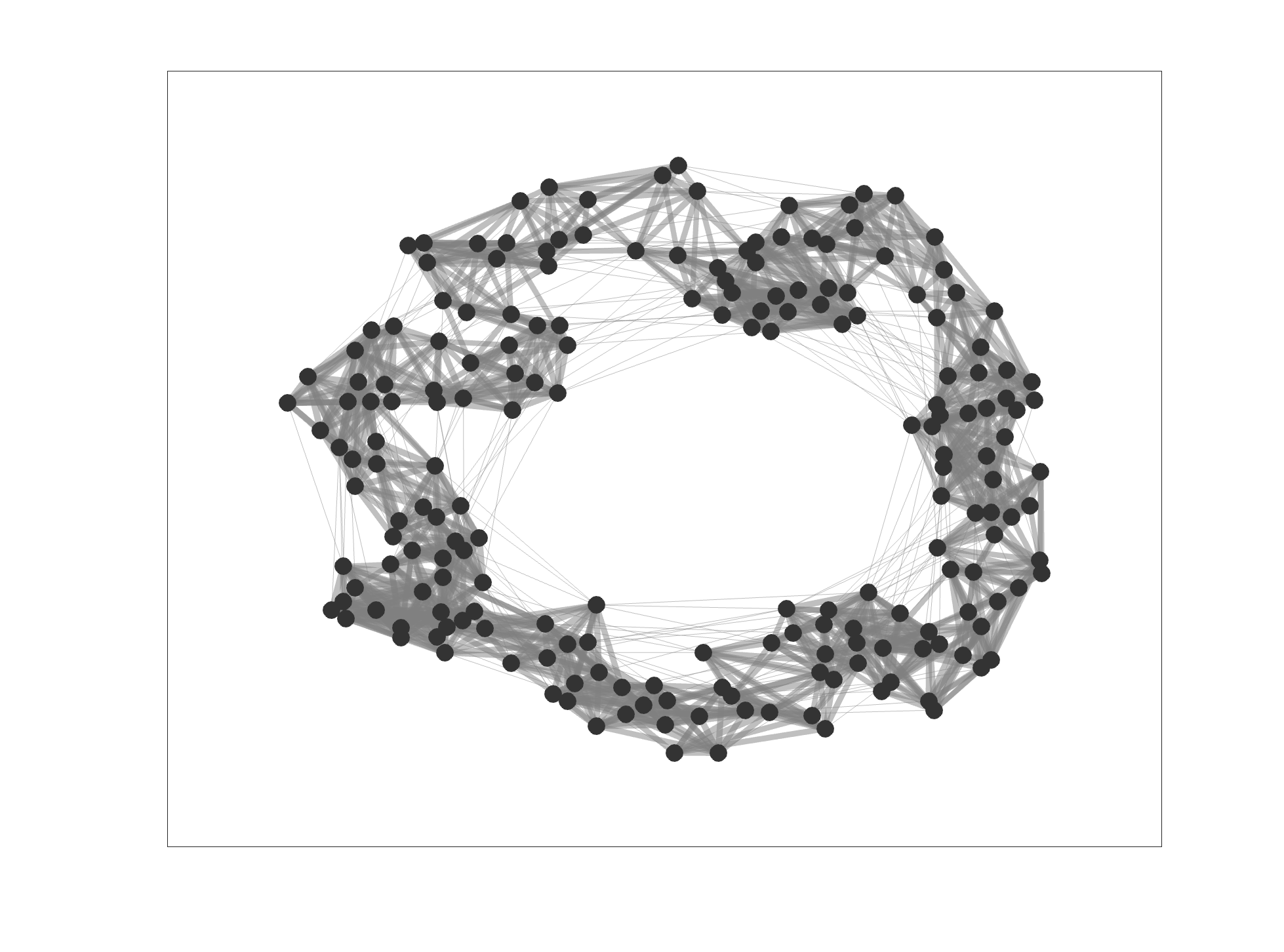}
    \caption{Network of related activities derived from the parametrization presented in figure \ref{fig:toeplitzcap}.}
    \label{fig:ringnet}
\end{figure}

But can we use this model to generate the network structure of the research space, which follows a ring instead of a core-periphery structure? Or do we need to radically change our assumptions to obtain that shape?\\

To generate a ring type network we can use Toeplitz-like matrices for the capability levels. A Toeplitz matrix is constant along each diagonal. By setting diagonals with decreasing values or $r_{c,b}$ and $q_{p,b}$ we can define correlations among subsets of related activities.\\

Here, we use a parametrization where we combine a symmetric Toeplitz circulant matrix and a random matrix by using proportions of ($\alpha$) and ($1-\alpha$). A circulant matrix is a particular type of Toeplitz matrix that has periodic boundary conditions. A symmetric circulant matrix can be constructed by starting from a row that is symmetric with respect to the center. Here, we generate these circulant matrices using linearly spaced probabilities for $r_c$ and $q_p$ that grow symmetrically from the center column of the first row. Figure~\ref{fig:toeplitzcap} shows an example of this parametrization for a model with 200 economies, 200 activities, and 200 capabilities. We note that in this model talking about higher and lower complexity economies is not a useful construct, since economies do not differ on their average capability level (they are all equal on average), but in which subset of capabilities they are specialized in.\\

Figure \ref{fig:ringnet} visualizes the network derived form this parametrization using the same method than before (minimum spanning tree, plus links that are one standard deviation above the average weight). The visualization shows a clear ring structure mimicking the one observed in networks involving research fields. The connectivity pattern of this network can be interpreted as research fields having a few related activities that share capabilities among them (e.g. capabilities are more re-deployable between molecular biology and biochemistry, than between polymer sciences and experimental psychology). This results in a network structure where each field is connected to a few neighbors.\\

\begin{figure}[htbp]
    \centering
    \includegraphics[width=0.8\textwidth]{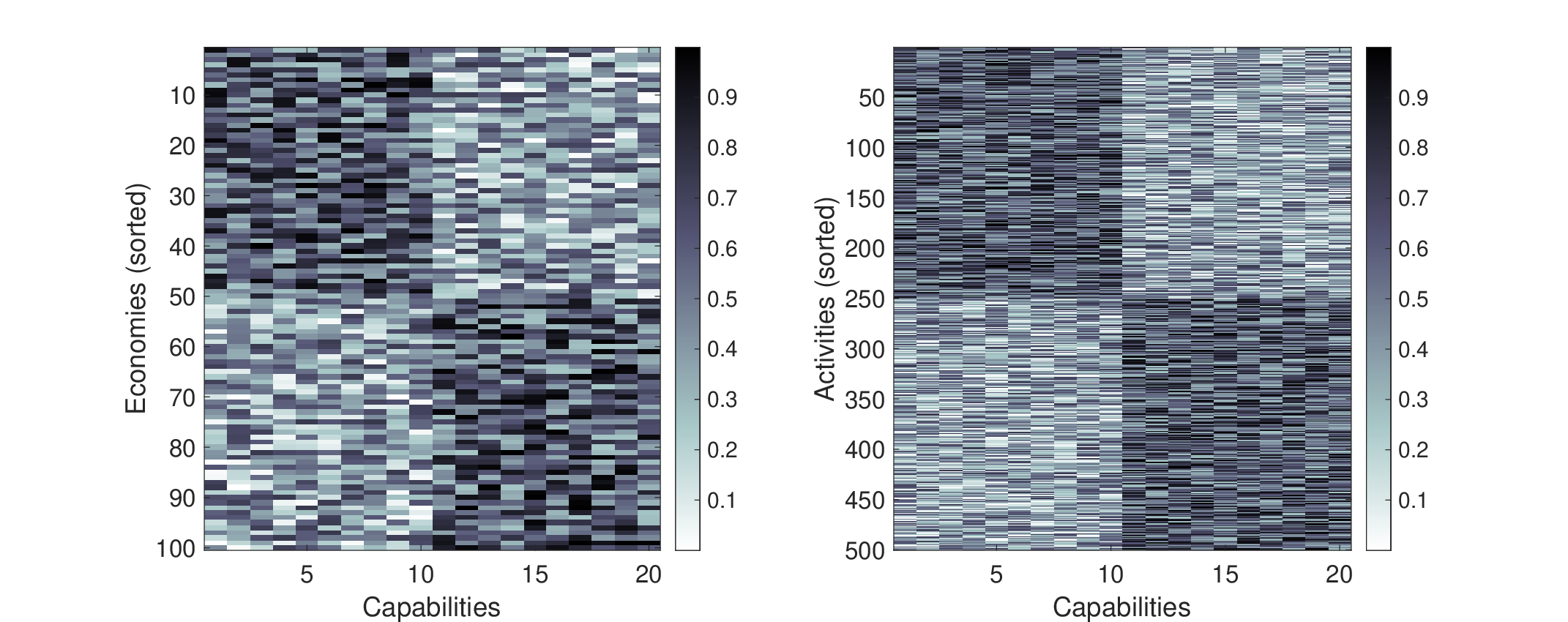}
    \caption{Parametrization of capability levels and requirements using a 25 percent of a block diagonal matrix and 75 percent of a random matrix.}
    \label{fig:dumbbellcap}
\end{figure}

\begin{figure}[htbp]
    \centering
    \includegraphics[width=0.7\textwidth]{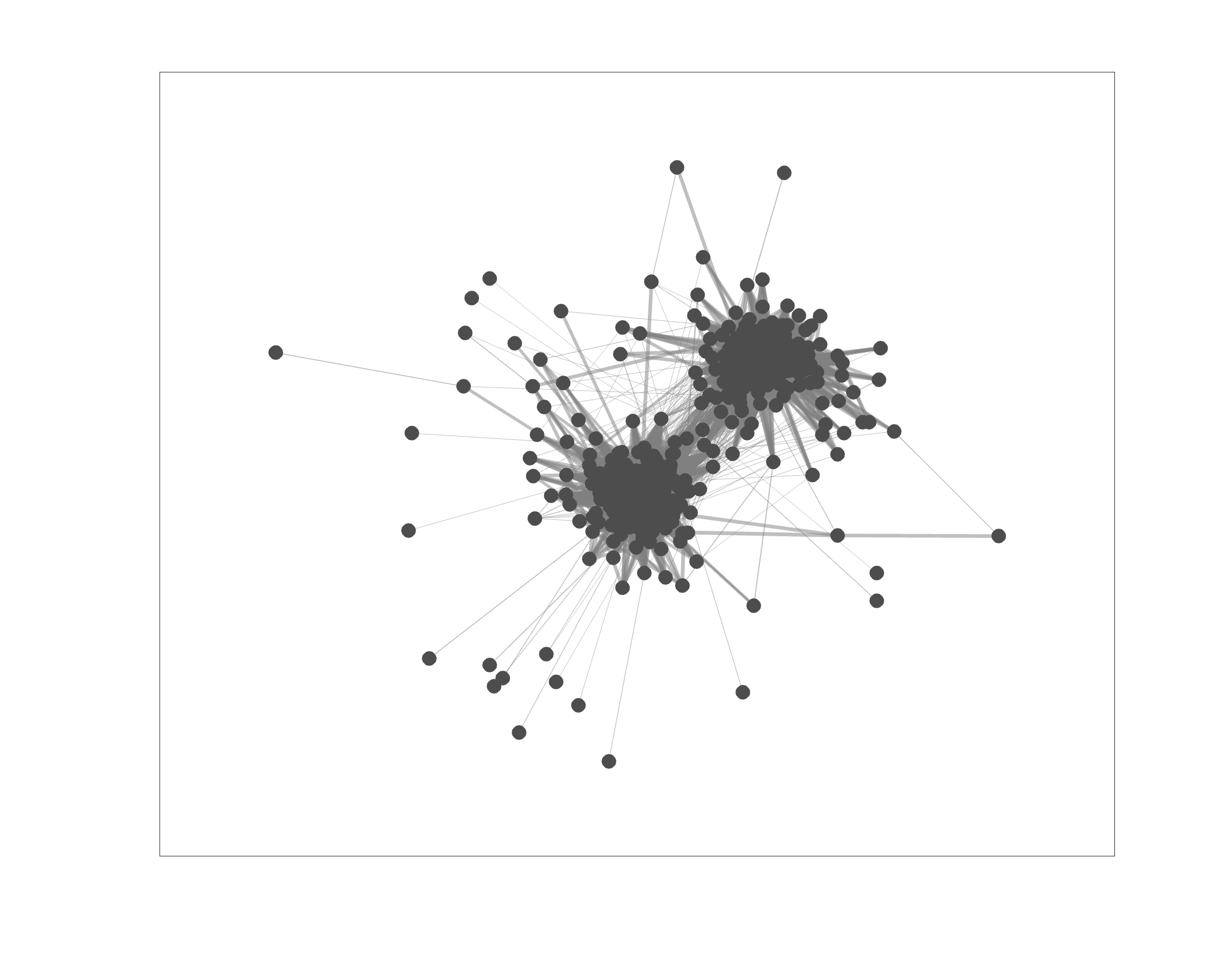}
    \caption{Network of related activities derived from the parametrization presented in figure \ref{fig:dumbbellcap}.}
    \label{fig:dumbbell}
\end{figure}

Finally, we use the same approach to model a ``dumbbell'' network, which is a network with two well-defined clusters, such as the one observed when connecting skills and occupations \cite{alabdulkareem_unpacking_2018}. Figures \ref{fig:dumbbellcap} and \ref{fig:dumbbell} show an example with 100 economies, 500 activities, and 20 capabilities. We note that obtaining this dumbbell structure requires a good level of mixing between the clusters, which can be achieved by setting the noise levels to be high enough so that some of the between cluster links are comparable in strength to the withing cluster links.\\

What is exciting about this general idea is that it provides us with an intuitive way to read capability levels from network structures. For instance, the core periphery-structure of the product space tells us that the capabilities associated with exporting products are correlated among economies, with high complexity economies like that of Singapore, Japan, or the United States, having developed a wide set of capabilities. The ring structure of the research space tells a different story. It is a story of specialization in a world of fine grained capabilities. Similarly, we can use this intuition to think about dumbbell structures, which can be modeled by assuming capability levels made of slightly overlapping blocks.\\ 

\section{Conclusion}
\label{sec:conclusion}

For long, economic complexity has attempted to study economic development using methods that are agnostic about the exact nature of factors of production. In this paper, we contributed to this goal by providing an analytical foundation for the economic complexity index ($ECI$) and showing that it can indeed be considered an estimate of the combined presence of undefined or unknown factors of production. 

In particular, we used a Kremer-Shockley production function as generative mapping from latent capability profiles to observable activity-level output and specialization patterns. The relevant question was not whether aggregate output can be represented by a smooth function of aggregate inputs, but what kinds of specialization matrices arise when locations differ in their capabilities and activities differ in their requirements. In this sense, the production function was a formal device for connecting unobserved capabilities to the observed matrices used in economic complexity analysis.\\

Indeed, for the single-capability model, we could derive the key eigenvector analytically and show that $ECI$ separates economies among those with an above- and below-average probability of having the capability. We then extended this result numerically to a multi-capability setting to show that $ECI$ is a monotonic estimator of an economy’s average capability level--even when a substantial share of the capabilities are randomly assigned. In the multi-capability model, $ECI$ is no longer a discrete measure separating low from high capability economies, but a monotonic transformation of the average capability level of an economy that recovers the first singular vector of the capability matrix $r_{cb}$. \\

These findings differentiate $ECI$ from measures of diversity, which peak for capability levels below the maximum (they are non-monotonic functions of $r_c$), and thus are non-ideal estimates of the complexity of an economy. These results validate $ECI$ as a measure of composition or complexity, since they show the eigenvector captures information about an economy having multiple capabilities, regardless of how these capabilities are defined. \\

Interestingly, our main result does not depend on assuming a stochastic model or a theory based on capabilities, since this idea can be easily generalized to models including factors that are specific to economies and activities. The key condition for the measure of complexity to work is for output to not be perfectly proportional to factor endowments. This condition can be achieved by simply shifting the production function by a constant to make it non-multiplicatively separable.\footnote{This mechanism is akin to the idea of symmetry-breaking in physics, since the shift removes symmetries of the function. For example $K_c^\gamma$ satisfies the scale-invariance symmetry $f(\lambda K)=\lambda^\gamma f(K)$, whereas $B+K_c^\gamma$ lacks this symmetry.} \\

What is also interesting is that the condition needed for our main result to hold comes from calculating the matrix of specialization $R_{cp}$. This is a key difference with previous attempts to connect economic complexity theory and empirics~\cite{hidalgo_building_2009,hausmann_network_2011,cristelli_measuring_2013} which jumped directly to the binary specialization matrix $M_{cp}$. That assumption results in a monotonic relationship between the number of activities an economy is specialized (its diversity) and its capability levels\footnote{That equation is provided in~\cite{hausmann_network_2011}.}, which is an uncomfortable result since we have known for a long time that measures of diversity fail to explain future economic growth like measures of complexity do~\cite{hidalgo_building_2009}. We now understand that calculating these specialization matrices is a key step, and that skipping this step in theoretical work results in a flawed connection between complexity and capability levels. This change not only uncovers a tight connection between the economic complexity index and the capability, but explains other findings, like that of Imbs and Wacziarg~\cite{imbs_stages_2003}, which says that economies diversify only until a certain point.\\

Our results also open questions about alternative measures of complexity. During the last fifteen years, many alternatives to the economic complexity index have been proposed, such as the Fitness index~\cite{tacchella_new_2012}, the Ability index~\cite{bustos_production_2022}, and many others~\cite{valverde-carbonell_rethinking_2025,sciarra_reconciling_2020,gnecco_machine_2022,atkin2021globalization,alqurtas_new_2018,ivanova_economic_2017,ivanova_measuring_2020,mcnerney_bridging_2023}. Since these indexes tend to exhibit strong correlations with $ECI$, our results provide a way to theoretically explore whether they are also monotonic functions of an economy's capability level, and to clarify which properties of the data transformation and the ranking procedure are essential for monotonicity. In particular, recent log-supermodularity results~\cite{schetter2022measure,yildirim_sorting_2021} highlight conditions under which eigenvector-based rankings are guaranteed to preserve the latent capability ordering in production functions in idealized, noise-free settings, but our findings emphasize why the empirical normalization steps embedded in $ECI$ matter. That is, real-world output data combine strong heterogeneity in the scale of economies and activities with measurement error and short-run shocks, and these features can break the strict log-supermodularity structure even when the underlying production component remains log-supermodular. In this paper we explored a few alternatives, showing that at least for the cases explored, $ECI$ is a justifiable methodology when output data involves heterogeneous units of observation and noise, precisely because its specialization-based normalization can remain informative about capability levels in settings where strict log-supermodularity of observed outcomes is not guaranteed.\\

Our work also speaks to the literature attempting to explain the economic complexity index. A key result in this literature is the idea that $ECI$ is a clustering algorithm~\cite{mealy_interpreting_2019,bottai_reinterpreting_2024}, separating economies into different groups. Our work is consistent with this idea and provides a theoretical interpretation for the clusters, as it shows that what $ECI$ is doing is providing a sigmoid function sorting economies into a high- and low-capability cluster. This sigmoid behavior is a well known feature of the second eigenvector or eigenfunction of diffusion maps, Fokker-Planck operators, and spectral clustering methods (e.g. see~\cite{nadler2005diffusion}). This provides an interesting link to the general idea of diffusion albeit in the context of a model of economic development, opening the door to the notion that these methods could be capturing a generalizable property of economic systems subject to spillovers.\\  

We also embedded this model in a short-run equilibrium framework to estimate wages, consumption, and prices. In this framework, wages increase with capability levels and prices are higher for products that demand more capabilities. The latter result is highly convex ($\sim1/(1-q)$), meaning that there is an important premium for producing high complexity products. Interestingly, prices do not strongly affect the specialization condition, meaning that they leave the connection between capability levels and economic complexity mostly unchanged.\footnote{We assume prices are the same across economies.}\\

Finally, we showed that the model can explain structural differences in networks of related activities, such as the product space and research space. By controlling the shape of the capability matrices, we were able to reproduce the core-periphery structure observed in the product space~\cite{hidalgo_product_2007}, the ring structure observed for scientific publications~\cite{borner_design_2012,guevara_research_2016}, and the dumbbell structure observed for networks of occupations and skills~\cite{alabdulkareem_unpacking_2018}.\\

Together, these findings help resolve a few long-standing tensions in the economic complexity literature. First, and most importantly, the disconnect between its empirical metrics and their theoretical underpinnings. Our findings show that $ECI$ is not an arbitrary or ad-hoc measure, but can be thought of as an estimator of an economy’s capabilities derived from its pattern of specialization. This is an interesting finding, since it provides a means to estimate the combined presence of factors or capabilities even when these cannot be identified.\\

Second, we use standard macroeconomic assumptions to estimate the wages and prices associated with this model, which help support the well known empirical fact that economies tend to converge to a level of income that is related to their economic complexity~\cite{hidalgo_building_2009,hausmann_atlas_2014,chavez_economic_2017,domini_patterns_2022,koch_economic_2021,stojkoski_impact_2016,stojkoski_relationship_2017,ourens_can_2012,poncet_economic_2013,stojkoski_multidimensional_2022,tacchella_new_2012,bustos_production_2022,atkin2021globalization,teixeira2022economic,hoeriyah2022economic,mao2021economic,basile_economic_2022,cardoso2023export,romero2021economic}.\\

And third, we provide theoretical underpinnings for the structure of the networks of related activities. Our findings show that the structure of these networks is a reflection of how capabilities are distributed across economies and activities.\\    

More broadly, our work helps clarify a field that had grown rapidly in its empirical scope while lacking a shared theoretical core. By grounding complexity metrics in production functions, and explaining the structure of networks of relatedness using a capability-based model, we offer a framework that not only explains the empirical robustness of $ECI$, but should also open new paths for integrating economic complexity ideas further into development economics and trade theory.\\

\bibliographystyle{abbrv}
\theendnotes
\bibliography{eci-bib-v2}

\section*{Acknowledgments}
This work owes a very special acknowledgment to Cristian Jara-Figueroa. In 2014, Cristian joined my (C\'esar's) group at the MIT Media Lab. During the first year of his Master's he worked on the mathematical theory of economic complexity producing an impressive internal manuscript with many results. Those results were never published, but stayed in my group. In 2025, while looking at Cristian's work, I realized we had made an important and simple mistake at the very beginning, which was to assume that the capability model was a model of $M_{cp}$ instead of a model of the output matrix $Y_{c,p}$. This required reworking everything, and motivated me to estimate the intermediate matrices in the model (such as $R_{cp}$). In my mind this work owes enormously to that effort by Cristian many years ago. We would also like to acknowledge comments by Johanness Wachs, Harasiddh Ganesh, and other members of the Center for Collective Learning. The section on prices and wages was motivated by a inspiring conversation with Jean Tirole.\\

We acknowledge the support of the European Union LearnData, GA no. 101086712 a.k.a. 101086712-LearnDataHORIZON-WIDERA-2022-TALENTS-01 (\url{https://cordis.europa.eu/project/id/101086712}), IAST funding from the French National Research Agency (ANR) under grant ANR-17-EURE-0010 (Investissements d'Avenir program), and the European Lighthouse of AI for Sustainability grant number 101120237-HORIZON-CL4-2022-HUMAN-02.

\end{document}